\documentclass[11pt]{article}

\usepackage{arxiv}

\usepackage[utf8]{inputenc} 
\usepackage[T1]{fontenc}    

\usepackage[hyphens]{url}

\usepackage{booktabs}       
\usepackage{amsfonts}       
\usepackage{nicefrac}       
\usepackage{microtype}      
\usepackage[square,numbers,sort&compress]{natbib}
\usepackage{listings}
\usepackage{caption}
\usepackage{float}
\usepackage{amssymb,amsmath}

\usepackage{longtable,booktabs}
\usepackage{graphicx}
\usepackage{subfig}
\usepackage{algorithm}
\usepackage{algorithmic}
\usepackage{array}

\title{An event-driven approach to serverless seismic imaging in the cloud}

\author{
  Philipp A.~Witte \\
  School of Computational Science and Engineering \\
  Georgia Institute of Technology \\
  Atlanta, GA 30308, USA \\
  \texttt{pwitte3@gatech.edu} \\
   \And
  Mathias~Louboutin \\
  School of Computational Science and Engineering \\
  Georgia Institute of Technology \\
  Atlanta, GA 30308, USA \\
  \texttt{mlouboutin3@gatech.edu} \\
   \And
  Henryk~Modzelewski \\
  Department of Earth, Ocean and Atmospheric Sciences \\
  University of British Columbia \\
  Vancouver, BC V6T 1Z4, Canada \\
  \texttt{hmodzelewski@eos.ubc.ca} \\
   \And
  Charles~Jones \\
  Osokey Ltd. \\
  Henley-on-Thames, RG9 1AY, UK \\
  \texttt{charles@osokey.com} \\
   \And
  James~Selvage \\
  Osokey Ltd. \\
  Henley-on-Thames, RG9 1AY, UK \\
  \texttt{james@osokey.com} \\
    \And
  Felix J.~Herrmann \\
  School of Computational Science and Engineering \\
  Georgia Institute of Technology \\
  Atlanta, GA 30308, USA \\
  \texttt{felix.herrmann@gatech.edu} \\
}

\newcommand\copyrighttext{%
  \footnotesize © 2019 IEEE.  Personal use of this material is permitted.  Permission from IEEE must be obtained for all other uses, in any current or future media, including reprinting/republishing this material for advertising or promotional purposes, creating new collective works, for resale or redistribution to servers or lists, or reuse of any copyrighted component of this work in other works.}
   
\begin{document}

\maketitle

\begin{abstract}
Adapting the cloud for high-performance computing (HPC) is a challenging task, as software for HPC applications hinges on fast network connections and is sensitive to hardware failures. Using cloud infrastructure to recreate conventional HPC clusters is therefore in many cases an infeasible solution for migrating HPC applications to the cloud. As an alternative to the generic lift and shift approach, we consider the specific application of seismic imaging and demonstrate a serverless and event-driven approach for running large-scale instances of this problem in the cloud. Instead of permanently running compute instances, our workflow is based on a serverless architecture with high throughput batch computing and event-driven computations, in which computational resources are only running as long as they are utilized. We demonstrate that this approach is very flexible and allows for resilient and nested levels of parallelization, including domain decomposition for solving the underlying partial differential equations. While the event-driven approach introduces some overhead as computational resources are repeatedly restarted, it inherently provides resilience to instance shut-downs and allows a significant reduction of cost by avoiding idle instances, thus making the cloud a viable alternative to on-premise clusters for large-scale seismic imaging.

\end{abstract}

\vspace*{2cm}
\copyrighttext
\clearpage

\section{Introduction}

Seismic imaging of the earth's subsurface is one of the most
computationally expensive applications in scientific computing, as
state-of-the-art imaging methods such as least-squares reverse time
migration (LS-RTM), require repeatedly solving a large number of forward
and adjoint wave equations during numerical optimization (e.g.
\cite{Valenciano2008, Dong2012, Zeng2014, Witte2019b}). Similar
to training neural networks, the gradient computations in seismic
imaging are based on backpropagation and require storage or
re-computations of the state variables (i.e.~of the forward modeled
wavefields). Due to the large computational cost of repeatedly modeling
wave propagation over many time steps using finite difference modeling,
seismic imaging requires access to high-performance computing (HPC)
clusters, but the high cost of acquiring and maintaining HPC cluster
makes this option only viable for a small number of major energy
companies \cite{Pgs2019, Mobile2019}. For this reason, cloud
computing has lately emerged as a possible alternative to on-premise HPC
clusters, offering many advantages such as no upfront costs, a
pay-as-you-go pricing model and theoretically unlimited scalability.
Outside of the HPC community, cloud computing is today widely used by
many companies for general purpose computing, data storage and analysis
or machine learning. Customers of cloud providers include major
companies such as General Electric (GE), Comcast, Shell or Netflix, with
the latter hosting their video streaming content on Amazon Web Services
(AWS) \cite{Customer2019}. Netflix' utilization of the cloud for
large-scale video streaming has acted as a driver for improving the
scalability of cloud tools such as object storage and event-driven
computations \cite{Cockroft2011}, which are not available on
conventional HPC environments and which we will subsequently adapt for
our purposes.

However, adapting the cloud for high-performance computing applications
such as seismic imaging, is not straight-forward, as numerous
investigations and case studies have shown that the cloud generally
cannot provide the same performance, low latency, high bandwidth and
mean time between failures (MTBF) as conventional HPC clusters. An early
performance analysis by Jackson \cite{jackson2010} of a range of typical NERSC
HPC applications on Amazon's Elastic Compute Cloud (EC2) found that, at
the time of the comparison, applications on EC2 ran between 2.7 to 50
times slower than on a comparable HPC system due to poor network
performance and that the latency was up to 400 times worse. A
performance analysis by \cite{iosup2011} using standard benchmark
suites such as the HPC challenge (HPCC) supports these observations,
finding that the performance on various cloud providers is in the order
of one magnitude worse than to comparable HPC clusters. Other
performance studies using standardized benchmarks suites, as well as
domain-specific applications, similarly conclude that poor network
performance severely limits the HPC capabilities of the cloud
\cite{garfinkel2007, napper2009, jackson2011, ramakrishnan2011, ramakrishnan2012, benedict2013, mehrotra2016}.

While communication and reliability are the strongest limiting factors
in the performance of HPC applications in the cloud, several investigations
\cite{gupta2011, sadooghi2017, kotas2018} point out that embarrassingly
parallel applications show in fact very good performance that is
comparable to (non-virtualized) HPC environments. Similarly, performance
tests on single cloud nodes and bare-metal instances using HPCC and
high-performance LINPACK benchmarks \cite{Dongarra2003}, demonstrate
good performance and scalability as well
\cite{rad2015b, mohammadi2018}. These findings underline that the \textit{lift
and shift approach} for porting HPC applications to the cloud is
unfavorable, as most HPC codes are based on highly synchronized message
passing (i.e.~MPI \cite{Gropp1999}) and rely on stable and fast network
connections, which are not (yet) available. On the other hand, compute
nodes and architectures offered by cloud computing are indeed comparable
to current supercomputing systems \cite{mohammadi2018} and the cloud
offers a range of novel technologies such as cloud object storage or
event-driven computations \cite{Lambda2019}. These technologies are not
available on traditional HPC systems and make it possible to address
computational bottlenecks of HPC in fundamentally new ways. Cloud object
storage, such as Amazon's Simple Storage Service (S3) \cite{S3AWS2019}
or Google Cloud Storage \cite{GoogleCloudStorage2019}, are based on the
distribution of files to physically separated data centers and thus
provide virtual unlimited scalability, as the storage system is not
constrained by the size and network capacity of a fixed number of servers
\cite{Barton2015}. Successfully porting HPC applications to the cloud
therefore requires a careful re-architecture of the corresponding codes
and software stacks to take advantage of these technologies, while
minimizing communication and idle times. This process is heavily
application dependent and requires the identification of how a specific
application can take advantage of specialized cloud services such as
serverless compute or high throughput batch processing to mitigate
resilience issues, avoid idle instances and thus minimize cost.

Based on these premises, we present a workflow for large-scale seismic
imaging on AWS, which does not rely on a conventional cluster of virtual
machines, but is instead based on a serverless visual workflow that
takes advantage of the mathematical properties of the seismic imaging
optimization problem \cite{Friedmann2017}. Similar to deep learning,
objective functions in seismic imaging consist of a sum of (convex)
misfit functions and iterations of the associated optimization
algorithms exhibit the structure of a MapReduce program
\cite{Dean2008}. The map part corresponds to computing the gradient of
each element in the sum and is embarrassingly parallel to compute, but
individual gradient computations are expensive as they involve solving
partial differential equations (PDEs). The reduce part corresponds to
the summation of the gradients and update of the model parameters and is
comparatively cheap to compute, but I/O intensive. Instead of performing
these steps on a cluster of permanently running compute instances, our
workflow is based on specialized AWS services such as AWS Batch and
Lambda functions, which are responsible for automatically launching 
and terminating the required computational resources \cite{Batch2019, 
Lambda2019}. EC2 instances are only running as long as they are utilized 
and are shut down automatically as soon as computations are finished, 
thus preventing instances from sitting idle. This stands in contrast to 
alternative MapReduce cloud services, such as Amazon's Elastic Map Reduce 
(EMR), which is based on Apache Hadoop and relies on a cluster of permanently 
running EC2 instances \cite{Hadoop2019, EMR2019}. In our approach,
expensive gradient computations are carried out by AWS Batch, a service
for processing embarrassingly parallel workloads, but with the
possibility of using (MPI-based) domain decomposition for individual
solutions of partial differential equations (PDEs). The cheaper gradient
summations are performed by Lambda functions, a service for serverless
computations, in which code is run in response to events, without the
need to manually provision computational resources \cite{Lambda2019}.

The following section provides an overview of the mathematical problem
that underlies seismic imaging and we identify possible characteristics
that can be taken advantage of to avoid the aforementioned shortcomings
of the cloud. In the subsequent section, we describe our seismic imaging
workflow, which has been developed specifically for AWS, but the
underlying services are available on other cloud platforms (Google
Compute Cloud, Azure) as well. We then present a performance analysis of
our workflow on a real-world seismic imaging application, using a
popular subsurface benchmark model \cite{Billette2005}. Apart from
conventional scaling tests, we also consider specific cloud metrics such
as resilience and cost, which, aside from the pure performance aspects
like scaling and time-to-solution, are important practical
considerations for HPC in the cloud. An early application of our
workflow is presented in \cite{Witte2019c}.

\section{Problem Overview}

Seismic imaging and parameter estimation are a set of computationally
challenging inverse problems with high practical importance, as they are
today widely used in the oil and gas (O\&G) industry for geophysical
exploration, as well as for monitoring geohazards. In the context of
exploration, seismic imaging can significantly increase the success rate
of drilling into reservoirs, thus reducing both cost and environmental
impact of resource exploration \cite{Advances2019}.

\begin{figure}[!tb]
\centering
\includegraphics[width=.60\hsize]{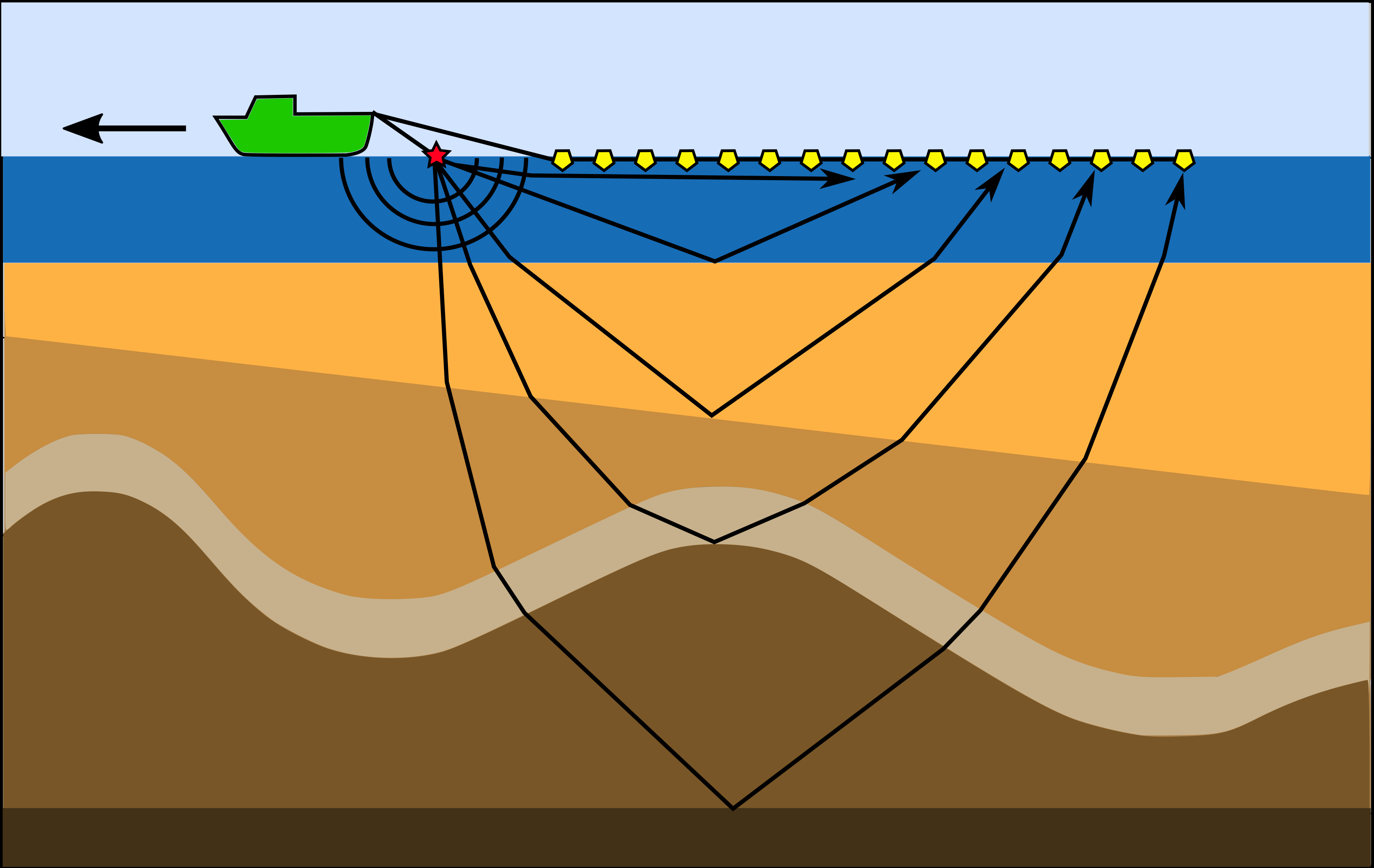}
\caption{A two-dimensional depiction of marine seismic data acquisition.
A vessel fires a seismic source and excites acoustic waves that travel
through the subsurface. Waves are reflected and refracted at geological
interfaces and travel back to the surface, where they are recorded by an
array of seismic receivers that are towed behind the vessel. The
receivers measure pressure changes in the water as a function of time
and receiver number for approximately 10 seconds, after which the
process is repeated. A typical seismic survey consists of several
thousand of these individual source experiments, during which the vessel
moves across the survey area.}\label{f0}
\end{figure}

Mathematically, seismic imaging and parameter estimation are
PDE-constrained optimization problems, that are typically expressed in
the following (unconstrained) form
\cite{Tarantola1984, Virieux2009}:
\begin{equation}
\begin{split}
    \underset{\mathbf{m}}{\operatorname{minimize}} \hspace{0.2cm} \Phi(\mathbf{m}) = \sum_{i=1}^{n_s} \frac{1}{2}||\mathcal{F}(\mathbf{m}, \mathbf{q}_i) - \mathbf{d}_i ||^2_2, \\
\end{split}
\label{objective}
\end{equation}
 where $\mathcal{F}(\mathbf{m}, \mathbf{q}_i)$ represents the solution
of the acoustic wave equation for a given set of model parameters
$\mathbf{m}$. The evaluation of this operator corresponds to modeling
seismic data for a given subsurface model (or image) $\mathbf{m}$ and a
known source function $\mathbf{q}_i$ by solving a wave equation using
time-domain finite-difference modeling. The vector $\mathbf{d}_i$
denotes the observed seismic measurements at the $i^\text{th}$ location
of the seismic source, which is moved along the surface within the
survey area (Figure~\ref{f0}). In essence, the goal of seismic inversion is
to find a set of model parameters $\mathbf{m}$, such that the
numerically modeled data matches the observed data from the seismic
survey. The total number of individual source experiments $n_s$ for
realistic surveys, i.e.~the number of PDEs that have to solved for each
evaluation of $\Phi(\mathbf{m})$, is quite large and lies in the range
of $10^3$ for 2D surveys and $10^5$ for 3D surveys.

Seismic inverse problems of this form are typically solved with
gradient-based optimization algorithms such as (stochastic) gradient
descent, (Gauss-) Newton methods, sparsity-promoting minimization or
constrained optimization (e.g. \cite{Pratt1999, peters2019}) and
therefore involve computing the gradient of Equation~\ref{objective} for
all or a subset of indices $i$. The gradient of the objective function
is given by:
\begin{equation}
    \mathbf{g} = \sum_{i=1}^{n_s} \mathbf{J}^\top \Big( \mathcal{F}(\mathbf{m}, \mathbf{q}_i) - \mathbf{d}_i \Big),
\label{gradient}
\end{equation}
 where the linear operator
$\mathbf{J} = \frac{\partial \mathcal{F}(\mathbf{m}, \mathbf{q}_i)}{\partial \mathbf{m}}$
is the partial derivative of the forward modeling operator with respect
to the model parameters $\mathbf{m}$ and $\top$ denotes the matrix
transpose. Both the objective function, as well as the gradient exhibit
a sum structure over the source indices and are embarrassingly parallel
to compute. Evaluating the objective function and computing the gradient
are therefore instances of a MapReduce program \cite{Dean2008}, as they
involve the parallel computation and subsequent summation of elements of
the sum. However, computing the gradient for a single index $i$ involves
solving two PDEs, namely a forward wave equation and an adjoint
(linearized) wave equation (denoted as a multiplication with
$\mathbf{J}^\top$). For realistically sized 3D problems, the discretized
model in which wave propagation is modeled has up to $10^9$ variables
and modeling has to be performed for several thousand time steps. The
number of time steps is determined by the time stepping interval and
depends on the wave speed and the temporal frequency of the data and
increases significantly as these properties change \cite{Courant1967}.
The observed seismic data $\mathbf{d}_i$ ($i=1,...,n_s$) is typically in
the range of several terabytes and a single element of the data (a
seismic \emph{shot record}) ranges from several mega- to gigabytes.

The problem structure of equation~\ref{objective} is very similar to
deep learning and the parallels between convolutional neural networks
and PDEs have lately attracted strong attention \cite{ruthotto2018}. As
in deep learning, computing the gradient of the objective function
(equation~\ref{gradient}) is based on backpropagation and in principle
requires storing the state variables of the forward problem. However, in
any realistic setting the wavefields are too big to be stored in memory
and therefore need to be written to secondary storage devices or
recomputed from a subset of checkpoints \cite{Griewank2000}.
Alternatively, domain decomposition can be used to reduce the domain
size per compute node such that the forward wavefields fit in memory, or
time-to frequency conversion methods can be employed to compute
gradients in the frequency domain \cite{Furse1998, Witte2019b}. In
either case, computing the gradient for a given index $i$ is expensive
both in terms of necessary floating point operations, memory and IO and
requires highly optimized finite-difference modeling codes for solving
the underlying wave equations. Typical computation times of a single
(3D-domain) gradient $\mathbf{g}_i$ (i.e.~one element of the sum) are in the
range of minutes to hours, depending on the domain size and the
complexity of the wave simulator, and the computations have to be
carried out for a large number of source locations and iterations.

The high computational cost of seismic modeling, in combination with the
complexity of implementing optimization algorithms to solve
equation~\ref{objective}, leads to enormously complex inversion codes,
which have to run efficiently on large-scale HPC clusters. A large
amount of effort goes into implementing fast and scalable wave equation
solvers \cite{abdelkhalek2009, weiss2013, Louboutin2018}, as well as
into frameworks for solving the associated inverse problem
\cite{Symes2011, ruthotto2016, daSilva2017, Witte2019}. Codes for
seismic inversion are typically based on message passing and use MPI to
parallelize the loop of the source indices (equation~\ref{objective}).
Furthermore, a nested parallelization is oftentimes used to apply
domain-decomposition or multi-threading to individual PDE solves. The
reliance of seismic inversion codes on MPI to implement an
embarrassingly parallel loop is disadvantageous in the cloud, where the
mean-time-between failures (MTBF) is much shorter than on HPC systems
\cite{jackson2010} and instances using spot pricing can be arbitrarily
shut down at any given time \cite{Spot2019}. Another important aspect
is that the computation time of individual gradients can vary
significantly and cause load imbalances and large idle times, which is
problematic in the cloud, where users are billed for running instances
by the second, regardless of whether the instances are in use or idle.
For these reasons, we present an alternative approach for seismic
imaging in the cloud based on batch processing and event-driven
computations.

\section{Event-driven seismic imaging on
AWS}\label{event-driven-seismic-imaging-on-aws}

\subsection{Workflow}\label{workflow}

Optimization algorithms for minimizing equation~\ref{objective}
essentially consists of three steps. First, the elements of the gradient
$\mathbf{g}_i$ are computed in parallel for all or a subset of indices
$i \in n_s$, which corresponds to the map part of a MapReduce program.
The number of indices for which the objective is evaluated defines the
batch size of the gradient. The subsequent reduce part consists of
summing these elements into a single array and using them to update the
unknown model/image according to the rule of the respective optimization
algorithm (Algorithm~\ref{alg:1}). Optimization algorithms that fit into
this general framework include variations of stochastic/full gradient
descent (GD), such as Nesterov's accelerated GD \cite{Nesterov2018} or
Adam \cite{Kingma2014}, as well as the nonlinear conjugate gradient
method \cite{Fletcher1964}, projected GD or iterative soft
thresholding \cite{Beck2009}. Conventionally, these algorithms
are implemented as a single program and the gradient computations for
seismic imaging are parallelized using message passing. Running
MPI-based programs of this structure in the cloud (specifically on AWS),
require that users request a set of EC2 instances and establish a
network connection between all workers \cite{EC22019}. Tools like
StarCluster \cite{Star2019} or AWS HPC \cite{Cluster2019} facilitate
the process of setting up a cluster and even allow adding or removing
instances to a running cluster. However, adding or dropping
instances/nodes during the execution of an MPI program is not easily
possible, so the number of instances has to stay constant during the
entire length of the program execution, which, in the case of seismic
inversion, can range from several days to weeks. This makes this
approach not only prone to resilience issues, but it can result in
significant cost overhead, if workloads are unevenly distributed and
instances are temporarily idle.

\begin{algorithm}[H]
\caption{Generic Algorithm Structure for Gradient-Based Minimization of
Equation~\ref{objective}, Using a Fixed Number of Iterations $n$. Many
Gradient-Based Algorithms Exhibit this Overall Map Reduce Structure,
Including Stochastic/Mini-Batch Gradient Descent (GD) with Various
Update Rules (e.g.~Nesterov, Adam), Projected GD, Iterative Soft
Thresholding or the Non-Linear Conjugate Gradient Method.}\label{alg:1}
\centering
\begin{algorithmic}[1]
\STATE Input:~batch~size~$n_b$,~max.~number~of~iterations~$n$,~step~size~$\alpha$,~initial~guess~$\mathbf{m}_1$
\FOR{$i=1$ to $n$}
\STATE Compute~gradients~$\mathbf{g}_i$,~$i=1, ..., n_b$~in~parallel
\STATE Sum~gradients:~$\mathbf{g} = \sum_{i=1}^{n_b} \mathbf{g}_i$
\STATE \text{Update~optimization~variable,~e.g.~using~SGD:}\newline ~$\mathbf{m}_{k+1} = \mathbf{m}_{k} - \alpha \mathbf{g}$
\ENDFOR
\end{algorithmic}
\end{algorithm}

\begin{figure}[!tb]
\centering
\includegraphics[width=0.60\hsize]{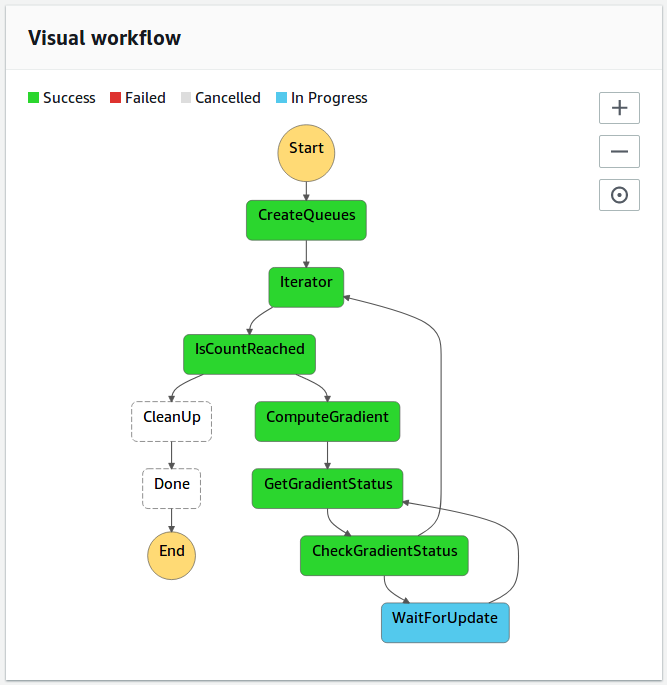}
\caption{A generic seismic imaging algorithm, expressed as a serverless
visual workflow using AWS Step Functions. The workflow consists as a
collection of states, which are used to implement an iterative
optimization loop. Each iteration involves computing the gradient of
Equation~\ref{objective} using AWS Batch, as well an updating the
optimization variable (i.e.~the seismic image).}\label{f1}
\end{figure}

Instead of implementing and running optimization algorithms for seismic
inverse problems as a single program that runs on a cluster of EC2
instances, we express the steps of a generic optimization algorithm
through AWS Step Functions (Figure~\ref{f1}) and deploy its individual
components through a range of specialized AWS services \cite{Step2019}.
Step functions allow the description of an algorithm as a collection of
states and their relationship to each other using the JavaScript Object
Notation (JSON). From the JSON definition of a workflow, AWS renders an
interactive visual workflow in the web browser, as shown in
Figure~\ref{f1}. For our purpose, we use Step Functions to implement our
iterative loop \cite{Loop2018}, during which we compute and sum the
gradients, and use them to update the seismic image. We choose Step
Functions to express our algorithm, as they allow composing different
AWS Services such as AWS Batch and Lambda functions into a single
workflow, thus making it possible to leverage preexisiting AWS services
and to combine them into a single application. Another important aspect
of Step Functions is that the execution of the workflow itself is
managed by AWS and does not require running any EC2 instances, which is
why we refer to this approach as \emph{serverless}. During execution
time, AWS automatically progresses the workflow from one state to the
next and users are only billed for transitions between states, but the
cost is negligible compared to the cost of running EC2 instances
(0.025\$ per 1,000 state transitions).

States can be simple if-statements such as the
\texttt{IsCountReached} state, which keeps track of the iteration number
and terminates the workflow after a specified number of iterations, but
states can also be used invoke other AWS services. Specifically, states
can be used to invoke AWS Lambda functions to carry out serverless
computations. Lambda functions allow users to run code in response to
events, such as invocations through AWS Step Functions, and automatically
assign the required amount of computational resources to run the code.
Billing is based on the execution time of the code and the amount of
used memory. Compared to EC2 instances, Lambda functions have a much
shorter startup time in the range of milliseconds rather than minutes,
but they are limited to 3 GB of memory and an execution time of 15
minutes. As such, Lambda functions themselves are not suitable for
carrying out the gradient computations, but they can be used to manage
other AWS services. In our workflow, we use Lambda functions invoked by
the \texttt{ComputeGradient} state (Figure~\ref{f1}) to launch AWS Batch
jobs for computing the gradients. During the gradient computation, which
can take up to several hours, the Step Functions check in a user-defined
interval if the full gradient has been computed, before advancing the
workflow to the next state. The \texttt{WaitForGradient} state pauses
the workflow for a specified amount of time, during which no additional
computational resources are running other than the AWS Batch job itself.

\subsection{Computing the gradient}\label{computing-the-gradient}

The gradient computations (equation~\ref{gradient}) are the major
workload of seismic inversion, as they involve solving forward and
adjoint wave equations, but the embarrassingly parallel structure of the
problem lends itself to high-throughput batch computing. On AWS,
embarrassingly parallel workloads can be processed with AWS Batch, a
service for scheduling and running parallel containerized workloads on
EC2 instances \cite{Batch2019}. Parallel workloads, such as computing a
gradient of a given batch size, are submitted to a batch queue and AWS
Batch automatically launches the required EC2 instances to process the
workload from the queue. Each job from the queue runs on an individual
instance or set of instances, with no communication being possible
between individual jobs.

\begin{figure}[!tb]
\centering
\includegraphics[width=0.600\hsize]{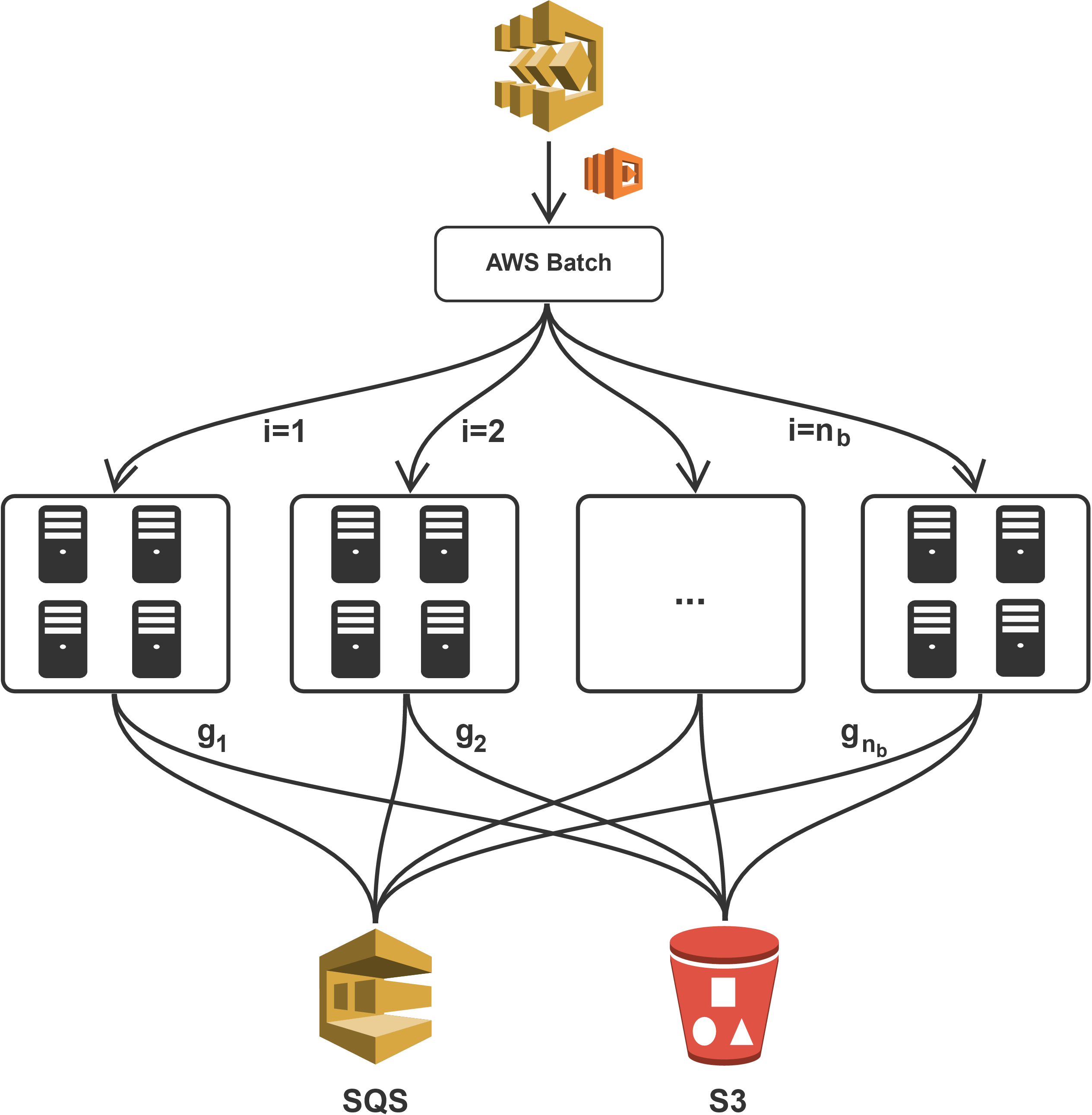}
\caption{The gradients of the LS-RTM objective function are computed as
an embarrassingly parallel workload using AWS Batch. This process is
automatically invoked by the AWS Step Functions (Figure~\ref{f1}) during
each iteration of the workflow. The gradients of individual source
locations are computed as separate jobs on either a single or multiple
EC2 instances. Communication is only possible between instances of a
single job, but not between separate jobs. The resulting gradients are
saved in S3 and the respective object names are sent to an SQS queue to
invoke the gradient summation.}\label{f2}
\end{figure}

In our workflow, we use the Lambda function invoked by the
\texttt{ComputeGradient} state (Figure~\ref{f1}) to submit the gradient
computations to an AWS Batch queue. Each element of the gradient
$\mathbf{g}_i$ corresponds to an individual job in the queue and is run
by AWS Batch as a separate Docker container \cite{Docker2019}. Every
container computes the gradient for its respective source index $i$ and
writes its resulting gradient to an S3 bucket (Figure~\ref{f2}),
Amazon's cloud object storage system \cite{S3AWS2019}. The gradients
computed by our workflow are one-dimensional numpy arrays of the size of
the vectorized seismic image and are stored in S3 as so-called objects
\cite{Numpy2011}. Once an individual gradient $\mathbf{g}_i$ has been
computed, the underlying EC2 instance is shut down automatically by AWS
Batch, thus preventing EC2 instances from idling. Since no communication
between jobs is possible, the summation of the individual gradients is
implemented separately using AWS Lambda functions. For this purpose,
each jobs also sends its S3 object identifier to a message queue (SQS)
\cite{SQS2019}, which automatically invokes the reduction stage
(Figure~\ref{f3}). For the gradient computations, each worker has to
download the observed seismic data of its respective source index from
S3 and the resulting gradient has to be uploaded to S3 as well. The
bandwidth with which objects are up- and downloaded is only limited by
the network bandwidth of the EC2 instances and ranges from 10 to 100
Gbps \cite{EC2Types2019}. Notably, cloud object storage such as S3 has
no limit regarding the number of workers that can simultaneously read
and write objects, as data is (redundantly) distributed among physically
separated data centers, thus providing essentially unlimited IO
scalability \cite{S3AWS2019}.

AWS Batch runs jobs from its queue as separate containers on a set of
EC2 instances, so the source code of the application has to be prepared
as a Docker container. Containerization facilitates portability and has
the advantage that users have full control over managing dependencies
and packages. Our Docker image contains the code for solving acoustic
wave equations to compute gradients of a respective seismic source
location. Since this is the most computational intensive part of our
workflow, it is important that the wave equation solver is optimized for
performance, but is also implemented in a programming language that
allows interfacing other AWS services such as S3 or SQS. In our
workflow, we use the domain-specific language compiler called Devito for
implementing and solving the underlying wave equations using time-domain
finite-difference modeling \cite{Louboutin2018, Luporini2018}. Devito
is implemented in Python and provides an application programming
interface (API) for implementing forward and adjoint wave equations as
high-level symbolic expressions based on the SymPy package
\cite{Joyner2012}. During runtime, the Devito compiler applies a series
of performance optimizations to the symbolic operators, such as
reductions of the operation count, loop transformations, and
introduction of parallelism \cite{Luporini2018}. Devito then generates
optimized finite-difference stencil code in C from the symbolic Python
expressions and dynamically compiles and runs it. Devito supports both
multi-threading using OpenMP, as well as generating code for MPI-based
domain decomposition. Its high-level API allows expressing wave
equations of arbitrary stencil orders or various physical
representations without having to implement and optimize low-level
stencil codes by hand. Furthermore, Devito includes various
possibilities for backpropagation, such as optimal checkpointing or
on-the-fly Fourier transforms \cite{Kukreja2018, Witte2019c}. The
complexity of implementing highly optimized and parallel wave equation
solvers is therefore abstracted and vertically integrated into the AWS
workflow.

By default, AWS Batch runs the container of each job on a single EC2
instance, but recently AWS introduced the possibility to run multi-node
batch computing jobs \cite{Rad2018}. Thus, individual jobs from the
queue can be computed on a cluster of EC2 instances and the
corresponding Docker containers can communicate via the AWS network. As
for single-instance jobs, AWS Batch automatically requests and
terminates the EC2 instances on which the Docker containers are
deployed. In the context of seismic imaging and inversion, multi-node
batch jobs enable nested levels of parallelization, as we can use AWS
Batch to parallelize the sum of the source indices, while using
MPI-based domain decomposition and/or multi-threading for solving the
underlying wave equations. This provides a large amount of flexibility
in regard of the computational strategy for performing backpropagation
and how to address the storage of the state variables. AWS Batch allows
to scale horizontally, by increasing the number of EC2 instances of
multi-node jobs, but also enables vertical scaling by adding additional
cores and/or memory to single instances. In our performance analysis, we
compare and evaluate different strategies for computing gradients with
Devito regarding scaling, costs and turnaround time.

\subsection{Gradient reduction}\label{gradient-reduction}

Every computed gradient is written by its respective container to an S3
bucket, as no communication between individual jobs is possible. Even if
all gradients in the job queue are computed by AWS Batch in parallel at
the same time, we found that the computation time of individual
gradients typically varies considerably (up to 10 percent), due to
varying network performance or instance capacity. Furthermore, we found
that the startup time of the underlying EC2 instances itself is highly
variable as well, so jobs in the queue are usually not all started at
the same time. Gradients therefore arrive in the bucket over a large
time interval during the batch job. For the gradient reduction step,
i.e.~the summation of all gradients into a single array, we take
advantage of the varying time-to-solutions by implementing an
event-driven gradient summation using Lambda functions. In this
approach, the gradient summation is not performed by as single worker or
the master process who has to wait until all gradients have been
computed, but instead summations are carried out by Lambda functions in
response to gradients being written to S3. The event-driven summation is
therefore started as soon as the first two gradients have been computed.

\begin{figure}[!tb]
\centering
\includegraphics[width=0.600\hsize]{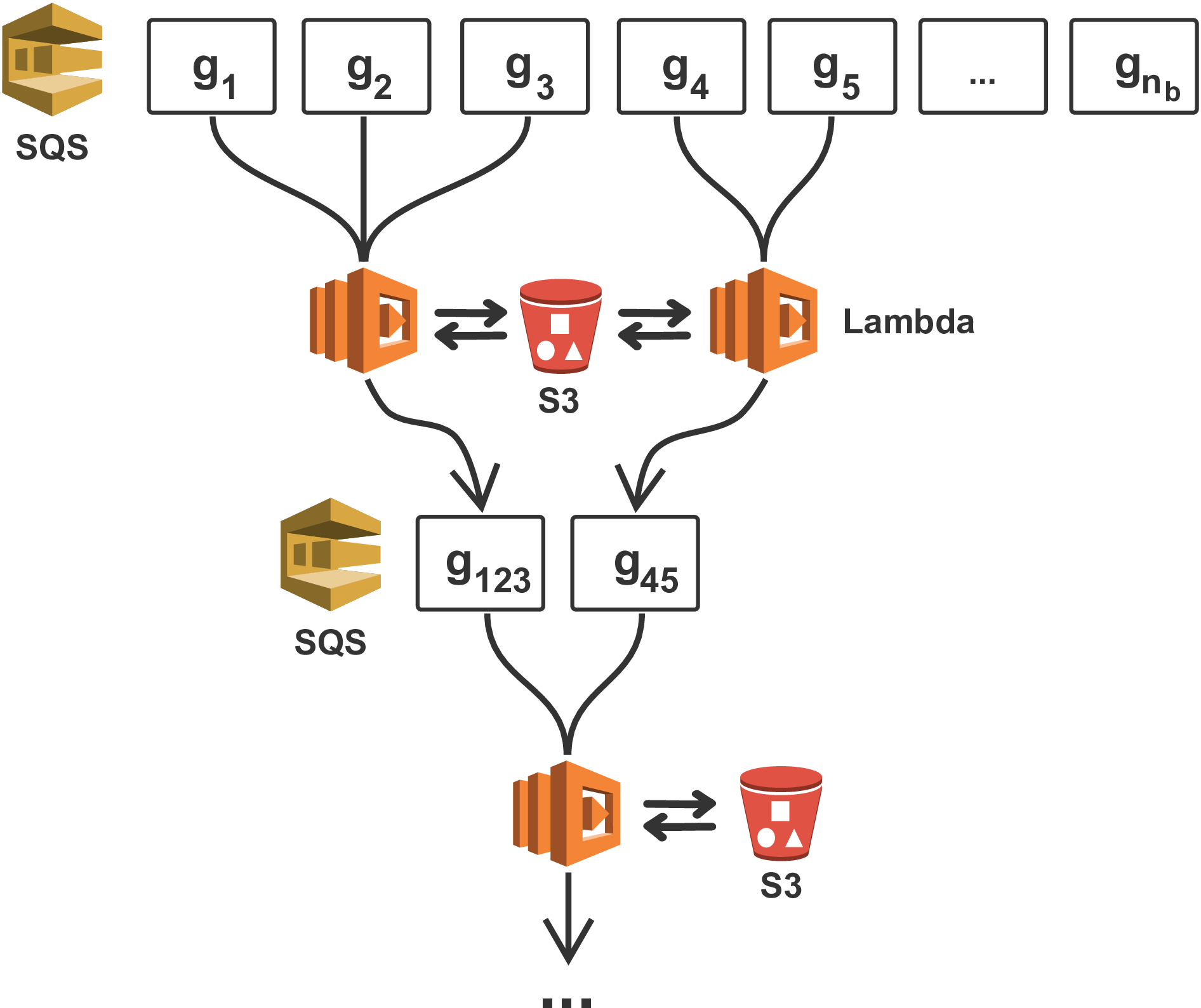}
\caption{Event-driven gradient summation using AWS Lambda functions. An
SQS message queue collects the object names of all gradients that are
currently stored in S3 and automatically invokes Lambda functions that
stream up to 10 files from S3. Each Lambda function sums the respective
gradients, writes the result back to S3 and sends the new object name to
the SQS queue. The process is repeated until all gradients have been
summed into a single S3 object. SQS has a guaranteed at-least-once
delivery of messages to ensure that no objects are lost in the
summation.}\label{f3}
\end{figure}

The event-driven gradient summation is automatically invoked through SQS
messages, which are sent by the AWS Batch workers that have completed
their computations and have saved their respective gradient to S3.
Before being shut down, every batch worker sends a message with the
corresponding S3 object name to an AWS SQS queue, in which all object
names are collected (Figure~\ref{f3}). Sending messages to SQS invokes
AWS Lambda functions that read up to 10 messages at a time from the
queue. Every invoked Lambda function that contains at least two
messages, i.e.~two object names, reads the corresponding arrays from S3,
sums them into a single array, and writes the array as a new object back
to S3. The new object name is sent to the SQS queue, while the previous
objects and objects names are removed from the queue and S3. The process
is repeated recursively until all $n_b$ gradients have been summed into
a single array, with $n_b$ being the batch size for which the gradient
is computed. The gradient summation is implemented in Python, which is
one of the languages supported by AWS Lambda \cite{Lambda2019}. SQS
guarantees that all messages are delivered at least once to the
subscribing Lambda functions, thus ensuring that no gradients are lost
in the summation process \cite{SQS2019}.

Since Lambda functions are limited to 3 GB of memory, it is not always
possible to read the full gradient objects from S3. Gradients that
exceed Lambda's available memory are therefore streamed from S3 using
appropriate buffer sizes and are re-uploaded to S3 using the
\texttt{multipart\_upload} functions of the S3 Python interface
\cite{Boto3}. As the execution time of Lambda functions is furthermore
limited to 15 minutes, the bandwidth of S3 is not sufficient to stream
and re-upload objects that exceed a certain size within a single Lambda
invocation. For this case, we include the possibility that the workers
of the AWS Batch job split the computed gradients into smaller chunks
that are saved separately in S3, with the respective objects names being
sent to multiple SQS queues. The gradient summation is then performed in
chunks by separate queues and Lambda functions. The
\texttt{CreateQueues} task of our Step Functions workflow
(Figure~\ref{f1}) automatically creates the required number of queues
before starting the optimization loop and the \texttt{CleanUp} state
removes them after the final iteration.

The advantage of the event-based gradient reduction is that that the
summation is executed asynchronously, as soon as at least two S3 objects
containing gradients are available, while other batch jobs are still
running. Therefore, by the time the last batch worker finishes the
computation of its respective gradient, all remaining gradients have
already been summed into a single object, or at least a small number of
objects. Furthermore, summing files of a single queue happens in
parallel (if enough messages are in the queue), as multiple Lambda
functions can be invoked at the same time. Furthermore, splitting the
gradients itself into chunks that are processed by separate queues leads
to an additional layer of parallelism. In comparison to a fixed cluster
of EC2 instances, the event-driven gradient summation using Lambda
function also takes advantage of the fact that the summation of arrays
is computationally considerably cheaper than solving wave equations and
therefore does not require to be carried out on the expensive EC2
instances used for the PDE solves.

\subsection{Variable update}\label{variable-update}

Once the gradients have been computed and summed into a single array
that is stored as an S3 object, the gradient is used to update the
optimization variables of equation~\ref{objective}, i.e.~the seismic
image or subsurface parameters such as velocity. Depending on the
specific objective function and optimization algorithm, this can range
from simple operations like multiplications with a scalars (gradient
descent) to more computational expensive operations such as sparsity
promotion or applying constraints \cite{Nocedal2006}. Updates that use
entry-wise operations only and are cheap to compute such as
multiplications with scalars or soft-thresholding, can be applied
directly by the Lambda functions in the final step of the gradient
summation. I.e. the Lambda function that sums the final two gradients,
also streams the optimization variable of the current iteration from S3,
uses the gradient to update it and directly writes the updated variable
back to S3.

Many algorithms require access to the full optimization variable and
gradient, such as Quasi-Newton methods and other algorithms that need to
compute gradient norms. In this case, the variable update is too
expensive and memory intensive to be carried out by Lambda functions and
has to be submitted to AWS Batch as a single job, which is then executed
on a larger EC2 instance. This can be accomplished by adding an extra
state such as \texttt{UpdateVariable} to our Step Functions workflow.
However, to keep matters simple, we only consider a simple stochastic
gradient descent example with a fixed step size in our performance
analysis, which is computed by the Lambda functions after summing the
final two gradients \cite{Bottou2010}. The \texttt{CheckGradientStatus}
state of our AWS Step Functions advances the workflow to the next iteration,
once the updated image (or summed gradient) has been written to S3. The
workflow shown in Figure~\ref{f1} terminates the optimization loop after
a predefined number of iterations (i.e.~epochs), but other termination
criteria based on gradient norms or function values are possible too and
can be realized by modifying the \texttt{IsCountReached} state. The
update of the optimization variable concludes a single iteration of our
workflow, whose performance we will now analyze in the subsequent
sections.

\section{Performance analysis}\label{performance-analysis}

\begin{figure}[!tb]
\centering
\includegraphics[width=1.000\hsize]{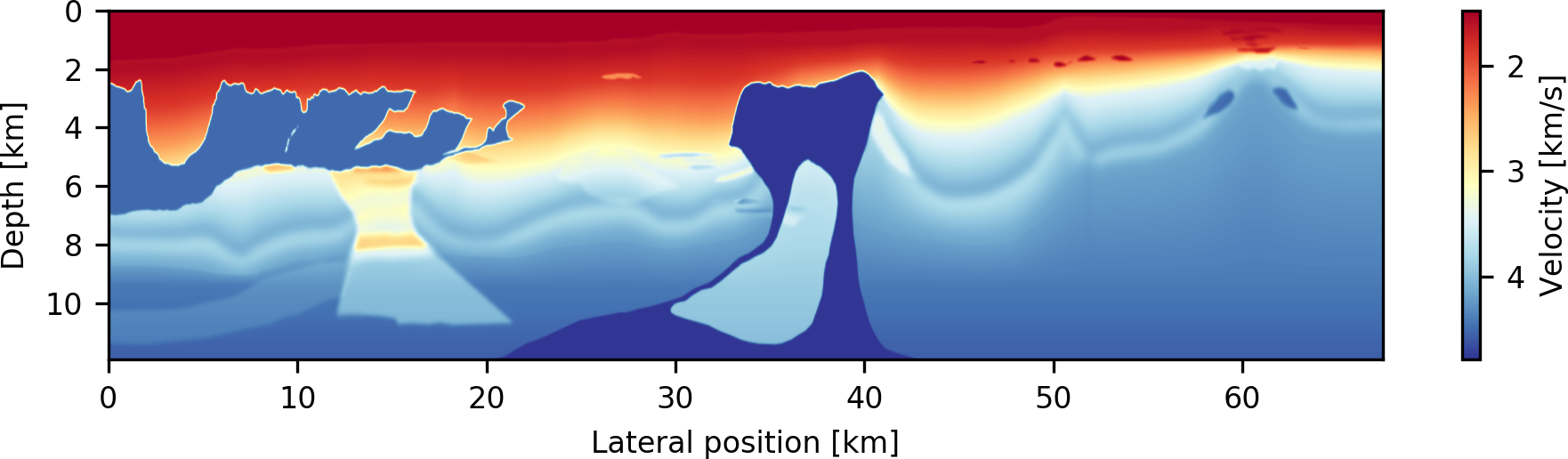}
\caption{The BP 2004 benchmark model, a 2D subsurface velocity model for
development and testing of algorithms for seismic imaging and parameter
estimation \cite{Billette2005}. This model and the corresponding
seismic data set are used in our performance analysis. The velocity
model and the unknown image have dimensions of $1,911 \times 10,789$
grid points, a total of 20.1 million unknown parameters.}\label{f4}
\end{figure}

In our performance analysis, we are interested in the performance of our
workflow on a real-world seismic imaging application regarding
scalability, cost and turn-around time, as well as the computational
benefits and overhead introduced by our event-driven approach. We
conduct our analysis on a popular 2D subsurface velocity model
(Figure~\ref{f4}), called the 2004 BP velocity estimation benchmark
model \cite{Billette2005}. This model was originally created for
analyzing seismic processing or inversion algorithms, but as the model
represents a typical large-scale 2D workload, we consider this model for
our following performance analysis. The seismic data set of this model
contains 1,348 seismic source locations and corresponding observations
$\mathbf{d}_i$ ($i=1, ..., 1,348$). The (unknown) seismic image has
dimensions of $1,911 \times 10,789$ grid points, i.e.~a total of almost
21 million parameters. An overview of all grid parameters, as well as
the dimensions of the seismic data are presented in Table~1 of
the Appendix.

\subsection{Weak scaling}\label{weak-scaling}

In our first performance test, we analyze the weak scaling behavior of
our workflow by varying the batch size (i.e.~the number of source
locations) for which the gradient of the LS-RTM objective function
(Equation~\ref{objective}) is computed. For this test, we perform a
single iteration of stochastic gradient descent (SGD) using our workflow
and measure the time-to-solution as a function of the batch size. The
workload per instance, i.e.~per parallel worker, is fixed to one
gradient. The total workload for a specified batch size is submitted to
AWS Batch as a so-called \emph{array job}, where each array entry
corresponds to a single gradient $\mathbf{g}_i$. AWS Batch launches one
EC2 instance per array entry (i.e.~per gradient), runs the respective
container on the instance and then terminates the instance afterwards.
If a sufficient amount of instances are available, AWS Batch will
theoretically launch all containers of the array job instantaneously and
run the full workload in parallel, but as we will see in the experiment,
this is in practice not necessarily the case.

In the experiment, we measure the time-to-solution for performing a
single iteration of our workflow, i.e.~one stochastic gradient descent
update. We exclude the setup time of the SQS queues, which is the first
step of our workflow (Figure~\ref{f1}), as this process only has to be
performed once, prior to the first iteration. Therefore, each run
involves the following steps:

\begin{enumerate}
\item
  A Lambda function submits the AWS Batch job for specified batch size
  $n_b$ (Figure~\ref{f2})
\item
  Compute gradients $\mathbf{g}_i$ ($i=1,...,n_b$) in parallel
  (Figure~\ref{f2})
\item
  Lambda functions sum the gradients (Figure~\ref{f3}): \newline
  $\mathbf{g} = \sum_{i=1}^{n_b} \mathbf{g}_i$
\item
  A Lambda function performs the SGD update of the image:
  $\mathbf{x} = \mathbf{x} - \alpha \mathbf{g}$
\end{enumerate}

We define the time-to-solution as the the time interval between the
submission of the AWS Batch job by a Lambda function (step 1) and the
time stamp of the S3 object containing the updated image (step 4). This
time interval represents a complete iteration of our workflow.

The computations of the gradients are performed on \texttt{m4.4xlarge}
instances and the number of threads per instance is fixed to 8, which is
the number of physical cores that is available on the instance. The
\texttt{m4} instance is a general purpose EC2 instance and we chose the
instance size (\texttt{4xlarge}) such that we are able to store the
wavefields for backpropagation in memory. The workload for each batch
worker consists of solving a forward wave equation to model the
predicted seismic data and an adjoint wave equation to backpropagate the
data residual and to compute the gradient. For this and all remaining
experiments, we use the acoustic isotropic wave equation with a second
order finite difference (FD) discretization in time and 8th order in
space. We model wave propagation for 12 seconds, which is the recording
length of the seismic data. The time stepping interval is given by the
Courant-Friedrichs-Lewy condition with 0.55 ms, resulting in 21,889
time steps. Since it is not possible for the waves to propagate through
the whole domain within this time interval, we restrict the modeling
grid to a size of $1,911 \times 4,001$ grid points around the current
source location. After modeling, each gradient is extended back to the
full model size ($1,911 \times 10,789$ grid points). A detailed
description of the setup parameters and utilized software and hardware
is provided in the appendix (Table~2). The dimensions of this
example represent a large-scale 2D example, but all components of our
workflow are agnostic to the number of physical dimensions and are
implemented for three-dimensional domains as well. The decision to limit
the examples to a 2D model was purely made from a financial viewpoint
and to make the results reproducible in a reasonable amount of time.

\begin{figure}[!tb]
\centering
\subfloat[\label{f5a}]{\includegraphics[width=0.460\hsize]{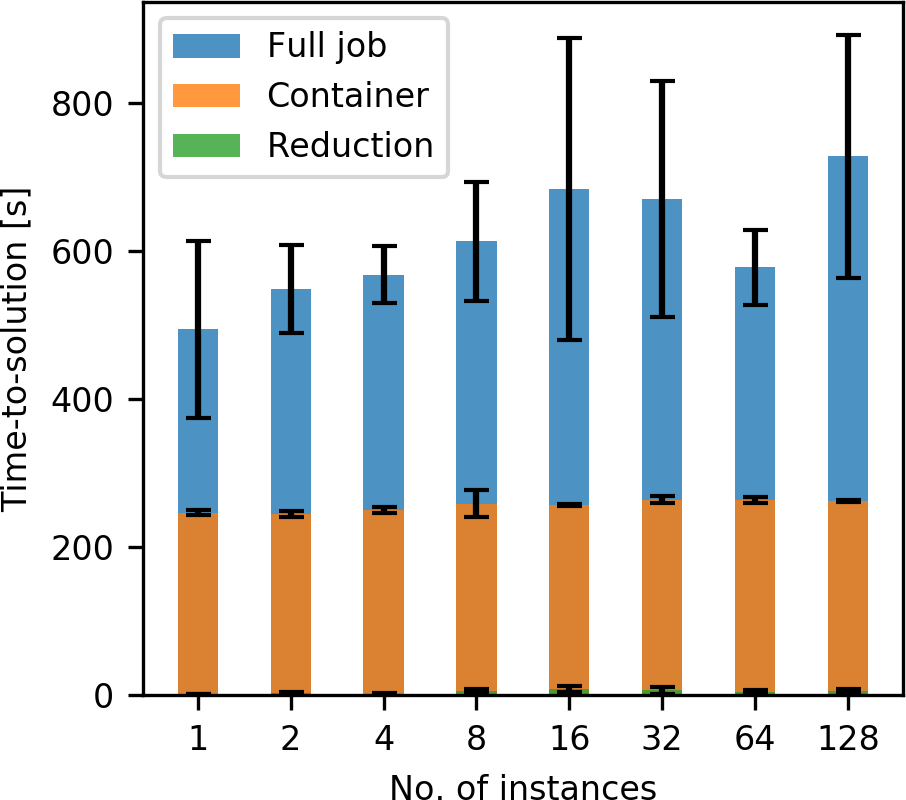}}
\hspace*{.6cm}
\subfloat[\label{f5b}]{\includegraphics[width=0.460\hsize]{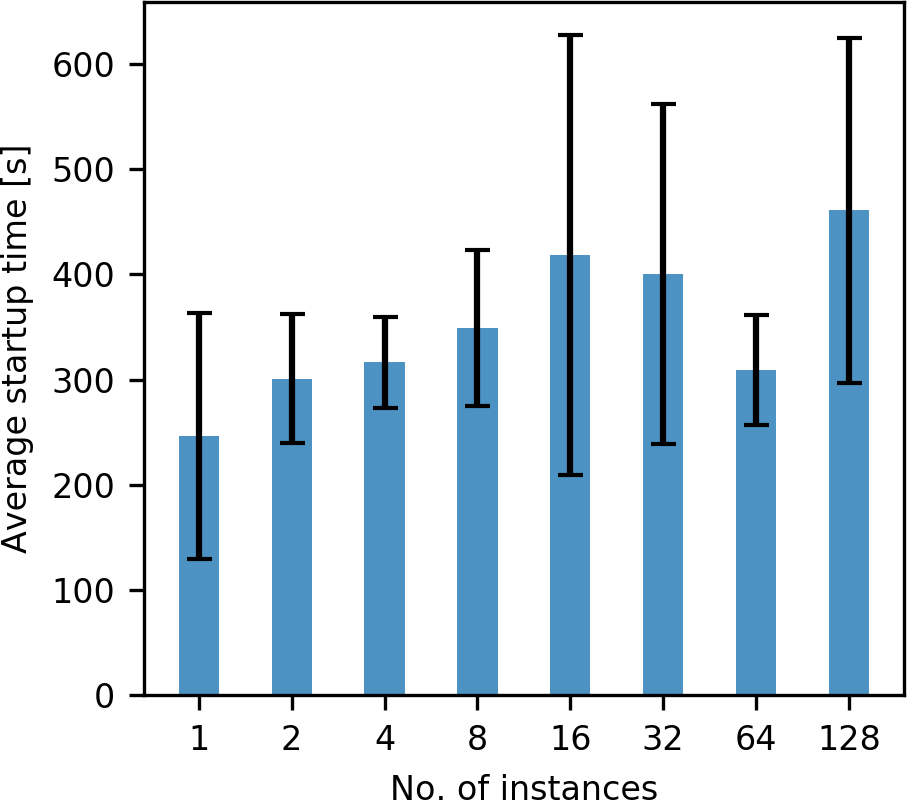}}
\\
\subfloat[\label{f5c}]{\includegraphics[width=0.485\hsize]{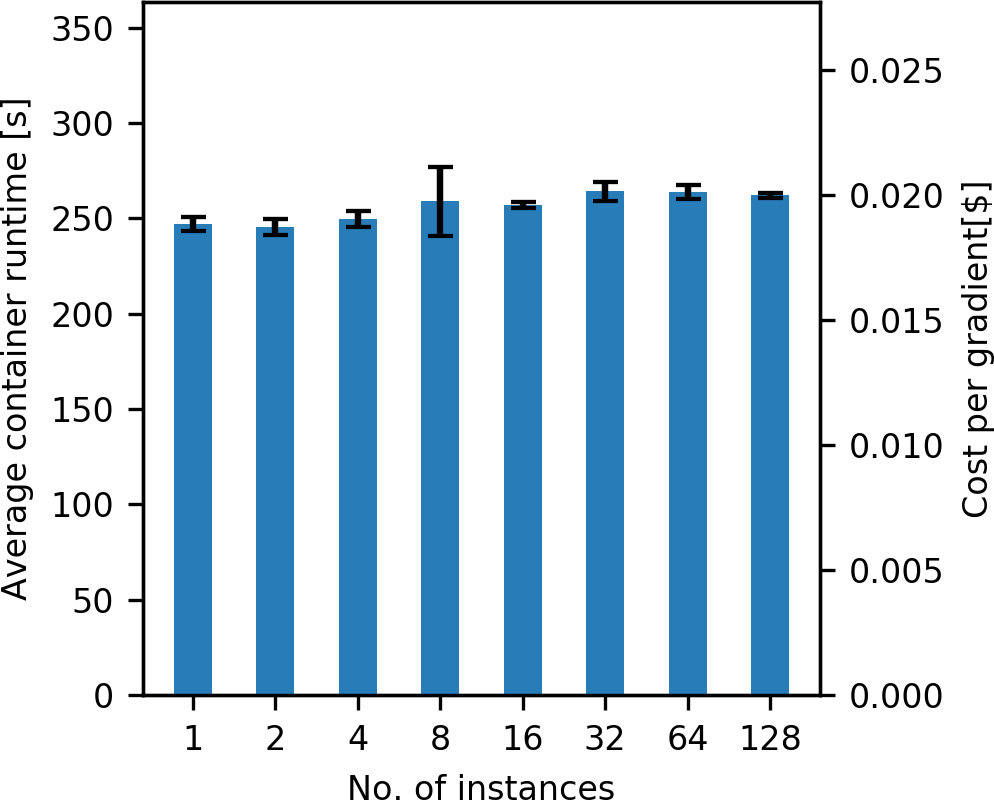}}
\hspace*{.6cm}
\subfloat[\label{f5d}]{\includegraphics[width=0.435\hsize]{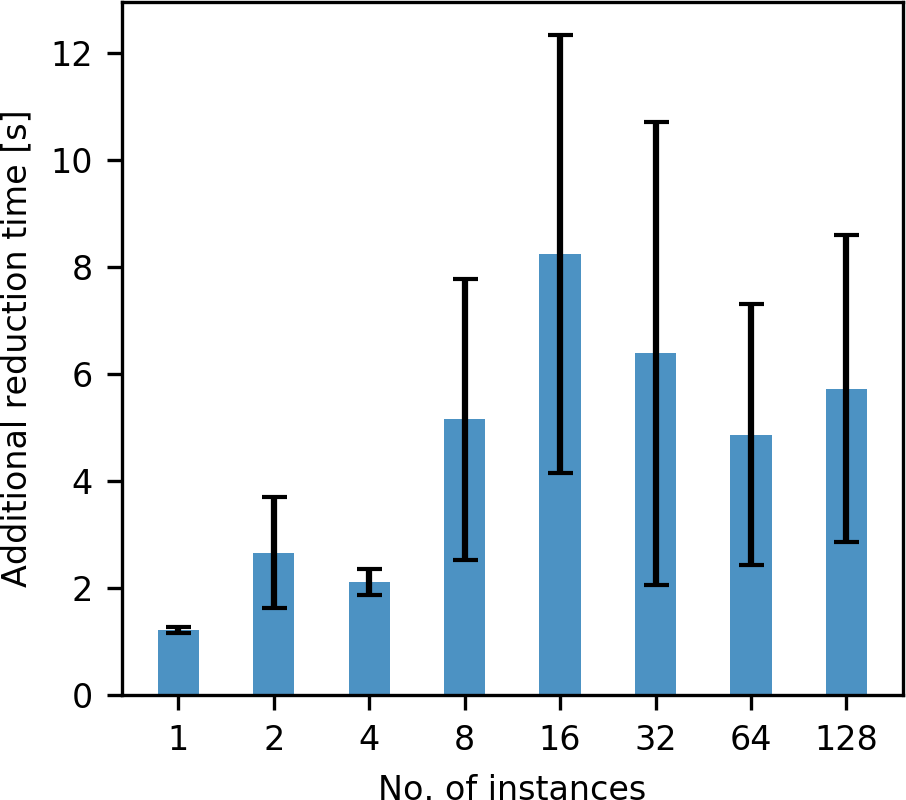}}
\caption{Weak scaling results for performing a single iteration of
stochastic gradient as a function of the batch size for which the
gradient is computed (a). The gradient is computed as an AWS Batch job
with an increasing number of parallel EC2 instances, while the gradient
summation and the variable update are performed by Lambda functions. The
total time-to-solution (a) consists of the average time it takes AWS
Batch to request and start the EC2 instances (b), the average runtime of
the containers (c) and the additional reduction time (d), i.e.~the time
difference between the final gradient of the respective batch and the
updated image. All timings are the arithmetic mean over three runs, with
the error bars representing the standard deviation.}\label{f5}
\end{figure}

The timings ranging from a batch size of 1 to 128 are displayed in
Figure~\ref{f5a}. The batch size corresponds to the number of parallel
EC2 instances on which the jobs are executed. The time-to-solution
consists of three components that make up the full runtime of each job:

\begin{enumerate}
\item
  The average time for AWS Batch to request and launch the EC2 instances
  and to start the Docker containers on those instances.
\item
  The runtime of the containers
\item
  The additional gradient reduction and image update time, which is
  given by the time interval between the termination of the AWS Batch
  job and the time stamp of the updated variable.
\end{enumerate}

The sum of these components makes up the time-to-solution as shown in
Figure~\ref{f5a} and each component is furthermore plotted separately in
Figures~\ref{f5b} to~\ref{f5d}. All timings are the arithmetic mean over
three individual runs and the standard deviation is indicated by the
error bars. The container runtimes of Figure~\ref{f5c} are the
arithmetic mean of the individual container runtimes on each instance
(varying from 1 to 128). The average container runtime is
proportional to the cost of computing one individual gradient and is
given by the container runtime times the price of the
\texttt{m4.4xlarge} instance, which was \$0.2748 per hour at the time
of testing. AWS Batch automatically launches and terminates the EC2
instance on which each gradient is computed and the user only pays for
utilized EC2 time. No extra charges occurs for AWS Batch itself,
i.e.~for scheduling and launching the batch job.

The timings indicate that the time-to-solution generally grows as the
batch size, and therefore the number of containers per job, increases
(Figure~\ref{f5a}). A close up inspection of the individual components
that make up the total time-to-solution shows that this is mostly due to
the increase of the startup time, i.e.~the average time it takes AWS
Batch to schedule and launch the EC2 instances of each job
(Figure~\ref{f5b}). We monitored the status of the EC2 instances during
the job execution and found that AWS Batch does generally not start all
instances of the array job at the same time, but instead in several
stages, over the course of 1 to 3 minutes. The exact startup time
depends on the batch size and therefore on the number of instances that
need to be launched, but also on the availability of the instance within
the AWS region. The combination of these factors lead to an increase of
the average startup time for an increasing batch size, but also to a
large variance of the startup time between individual runs.
Unfortunately, the user has no control over the startup time, but it is
important to consider that no cost is incurred during this time period,
as no EC2 instances are running while the individual containers remain
in the queue.

The average container runtime, i.e.~the average computation time of a
single gradient within the batch, is fairly stable as the batch size
increases (Figure~\ref{f5c}). This observation is consistent with the
fact that each container of an AWS Batch array job runs as an individual
Docker container and is therefore independent of the batch size. The
container runtime increases only slightly for larger batch sizes and we
observe a large variance in some of the container runtimes (specifically
for a batch size of 8). This variance stems from the fact that users do
not have exclusive access to the EC2 instances on which the containers
are deployed. Specifically, our containers run on \texttt{m4.4xlarge}
instances, which have 8 cores (16 virtual CPUs) and 64 GB of memory. In
practice, AWS deploys these instances on larger physical nodes and
multiple EC2 instances (of various users) can run on the same node. We
hypothesize that a larger batch size increases the chance of containers
being deployed to a compute node that runs at full capacity, thus
slightly increasing the average container runtime, as user do not have
exclusive access to the full network capacity or memory bandwidth.

\begin{figure}[!tb]
\centering
\includegraphics[width=1.000\hsize]{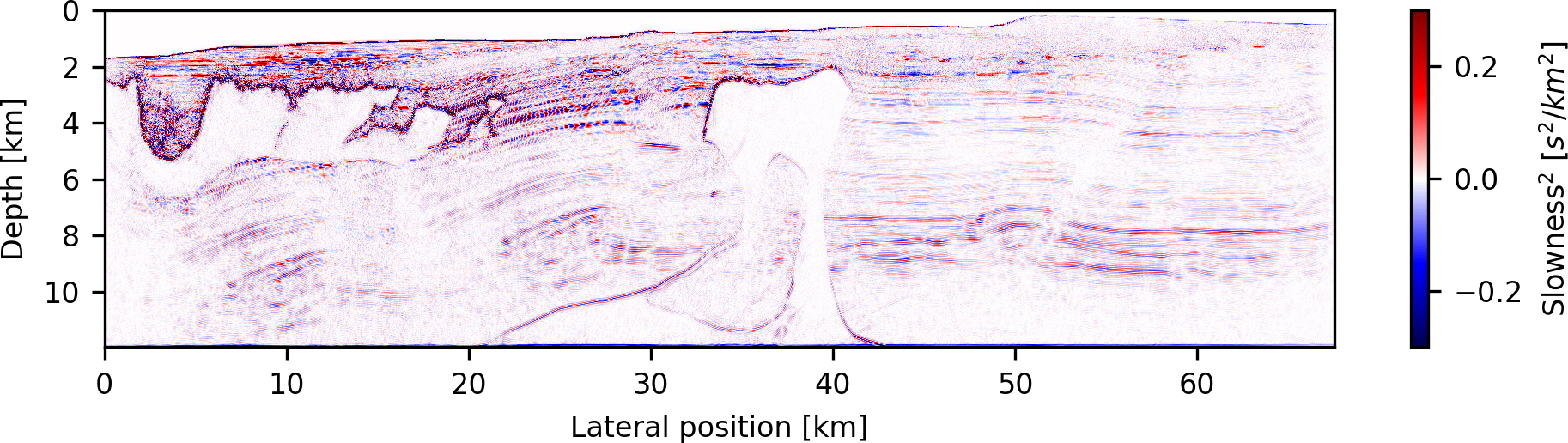}
\caption{Final seismic image after 30 iterations of stochastic gradient
descent and a batch size of 80, which corresponds to approximately two
passes through the data set (i.e.~two epochs).}\label{f6}
\end{figure}

Finally, we also observe an increase in the additional gradient
reduction time, i.e.~the interval between the S3 timestamps of the final
computed gradient $\mathbf{g}_i$ and the updated image $\mathbf{x}$. The
batch size corresponds to the number of gradients that have to be summed
before the gradient can be used to update the image.The event-driven
gradient reduction invokes the summation process as soon as the first
gradients are written to S3, so most gradients are already summed by the
time the final worker finishes its gradient computation. For the
event-driven gradient summation, the variance of the startup and
container runtime is therefore advantageous, as it allows the summation
to happen asynchronously. However, in our example, the time interval
between the first two gradients being written to S3 (thus invoking the
gradient reduction) and the final gradient being computed, does not
appear to be large enough to complete the summation of all gradients.
Specifically, we see an increase in the reduction time from a batch size
of 4 to 8, after which the additional reduction is mostly constant, but
again with a large variance. This variance is due to a non-deterministic
component of our event-based gradient summation, resulting from a
limitation of AWS Lambda. While users can specify a maximum number of
messages that Lambda functions read from an SQS queue, it is not
possible to force Lambda to read a minimum amount of two messages,
resulting in most Lambda functions reading only a single message
(i.e.~one object name) from the queue. Since we need at least two
messages to sum the corresponding gradients, we return the message to
the queue and wait for a Lambda invocation with more than one message.
The user has no control over this process and sometimes it takes several
attempts until a Lambda function with multiple messages is invoked. The
likelihood of this happening increases with a growing batch size, since
a larger number of gradients need to be summed, which explains the
increase of the reduction time and variance in Figure~\ref{f5d}.

Overall, the gradient summation and variable update finishes within a
few seconds after the last gradient is computed and the additional
reduction time is small compared to the full time-to-solution and to the
pure computation time of the gradients. In our example, the startup time
(Figure~\ref{f5b}) takes up the majority of the time-to-solution
(Figure~\ref{f5a}), as it lies in the range of a few minutes and is in
fact longer than the average container runtime of each worker
(Figure~\ref{f5c}). However, the startup time is independent of the
runtime of the containers, so the ratio of the startup time to the
container runtime improves as the workload per container increases.
Furthermore, the cost of the batch job only depends on the container
runtime and the batch size, but not on the startup time or reduction
time. The cost for summing the gradients is given by the cumulative
runtime of the Lambda functions, but is negligible compared to the EC2
cost for computing the gradients. At the time of the example, the cost
for Lambda functions was \$$2\cdot 10^{-7}$ per request and \$$1.6\cdot 10^{-5}$ per used
GB-second. Figure~\ref{f6} shows the final seismic image that is
obtained after running our workflow for 30 iterations and a batch size
of 80, which corresponds to 1.8 epochs. The source locations in each
iteration are chosen from a uniform random distribution and and after
the final iteration, each data sample (i.e.~seismic shot record) has
been, in expectation, used 1.8 times. In this example, every gradient
was computed by AWS Batch on a single instance and a fixed number of
threads, but in the subsequent section we analyze the scaling of runtime
and cost as a function of the number of cores and EC2 instances.
Furthermore, we will analyze in a subsequent example how the cost of
running the gradient computations with AWS Batch compares to performing
those computations on a fixed cluster of EC2 instances.

\subsection{Strong scaling}\label{strong-scaling}

In the following set of experiments, we analyze the strong scaling
behavior of our workflow for an individual gradient calculation, i.e.~a
gradient for a batch size of 1. For this, we consider a single gradient
computation using AWS Batch and measure the runtime as a function of
either the number of threads or the number of instances in the context
of MPI-based domain decomposition. In the first experiment, we evaluate
the vertical scaling behavior, i.e.~we run the gradient computation on a
single instance and vary the number of OpenMP threads. In contrast to
the weak scaling experiment, we model wave propagation in the full
domain ($1,911 \times 10,789$ grid points), to ensure that the
sub-domain of each worker is not too small when we use maximum number of
threads. The measured runtime is the sum of the kernel times spent for
solving the forward and the adjoint wave equation and therefore excludes
memory allocation time and code generation time.

Since AWS Batch runs all jobs as Docker containers, we compare the
runtimes with AWS Batch to running our application on a bare metal
instance, in which case we have direct access to the compute node and
run our code without any virtualization. All timings on AWS are
performed on a \texttt{r5.24xlarge} EC2 instance, which is a memory
optimized instance type that uses the Intel Xeon Platinum 8175M
architecture. The \texttt{24xlarge} instance has 96 virtual CPU cores
(48 physical cores on 2 sockets) and 768 GB of memory. Using the largest
possible instance of the \texttt{r5} class, ensures that our AWS Batch
job has exclusive access to the physical compute node. Bare metal
instances automatically give users exclusive access to the full node. We
also include the Optimum HPC cluster in our comparison, a small research
cluster at the University of British Columbia based on the Intel's Ivy
Bridge 2.8 GHz E5-2680v2 processor. Optimum has 2 CPUs per node and 10
cores per CPU.

\begin{figure}[!tb]
\centering
\subfloat[\label{f7a}]{\includegraphics[width=0.469\hsize]{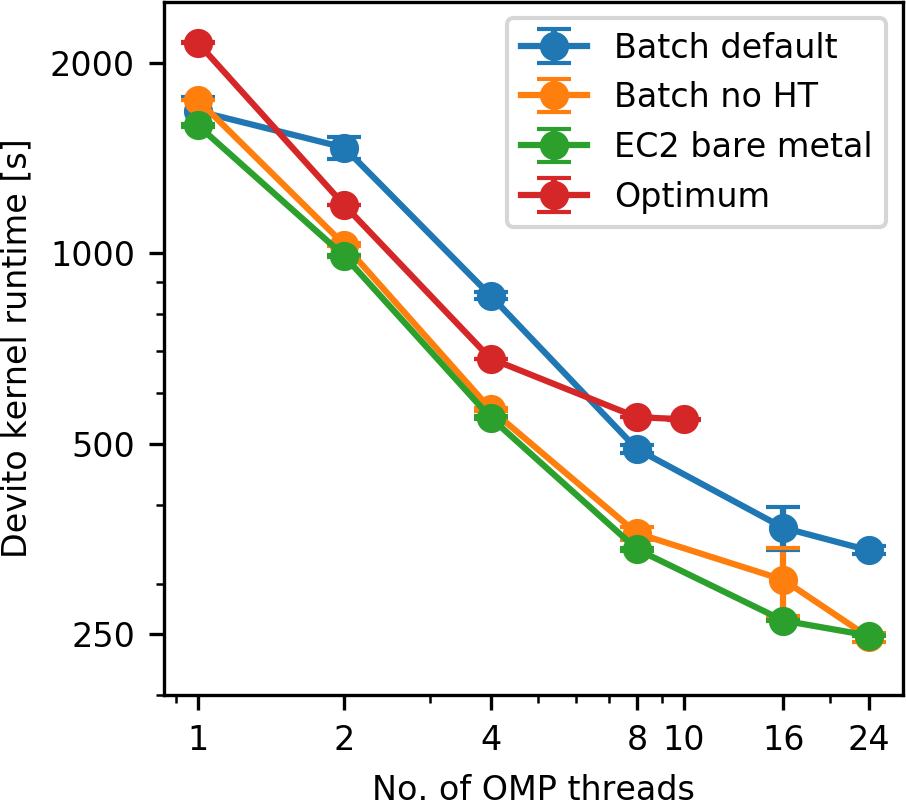}}
\hspace*{.6cm}
\subfloat[\label{f7b}]{\includegraphics[width=0.469\hsize]{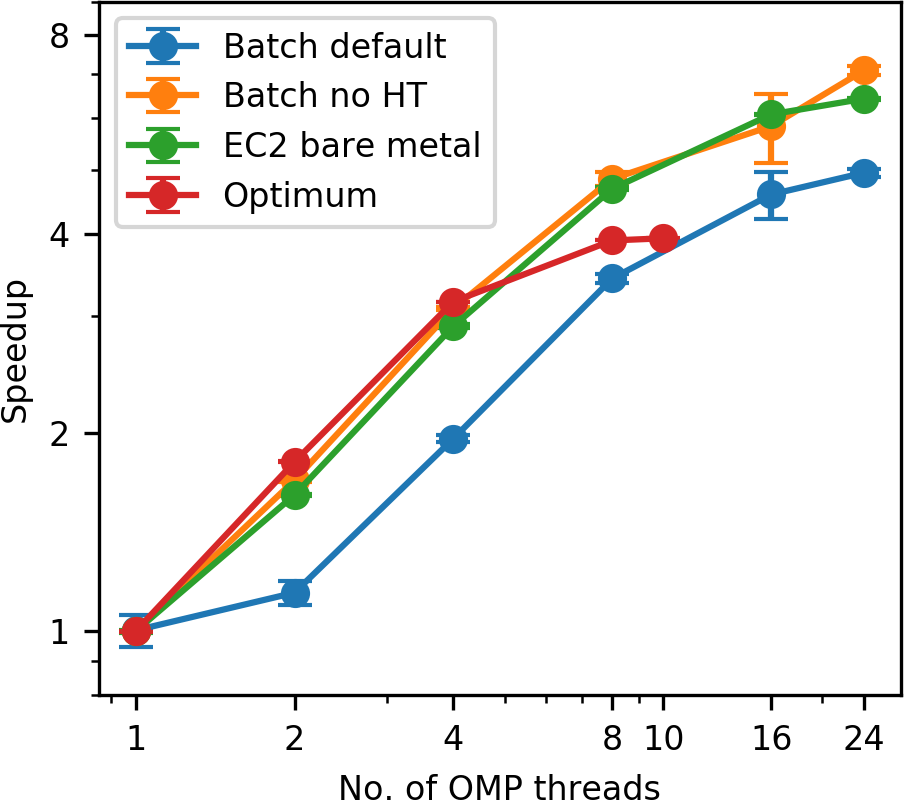}}
\caption{Strong scaling results for computing a single image gradient of
the BP model as a function of the number of threads. Figure (a) shows
the runtimes for AWS Batch with and without hyperthreading, as well as
the runtimes on the r5 bare metal instance, in which case no
containerization or virtualization is used. For reference, we also
provide the runtime on a compute node of an on-premise HPC cluster.
Figure (b) shows the corresponding speeds ups.}\label{f7}
\end{figure}

Figure~\ref{f7a} shows the comparison of the kernel runtimes on AWS and
Optimum and Figure~\ref{f7b} displays the corresponding speedups. As
expected, the \texttt{r5} bare metal instance shows the best scaling, as
it uses a newer architecture than Optimum and does not suffer from the
virtualization overhead of Docker. We noticed that AWS Batch in its
default mode uses hyperthreading (HT), even if we perform thread pinning
and instruct AWS Batch to use separate physical cores. As of now, the
only way to prevent AWS Batch from performing HT, is to modify the
Amazon Machine Image (AMI) of the corresponding AWS compute environment
and set the \texttt{nr\_cpus} parameter of the
\texttt{/etc/default/grub} file to the number of physical cores per
socket (i.e. 24). With HT disabled, the runtimes and speedups of AWS
Batch are very close to the timings on the bare-metal instances,
indicating that the overhead of Docker affects the runtimes and scaling
of our memory-intensive application only marginally, which matches the
findings of \cite{chung2016}.

Next, we analyze the horizontal strong scaling behavior of running our
application with AWS Batch. Once again, we consider the computation of
one single gradient, but this time we vary the number of EC2 instances
on which the underlying wave equations are solved. We would like to
emphasize that AWS Batch is used differently than in the weak scaling
experiment, where AWS Batch was used to parallelize the sum over source
locations. Multiple workloads (i.e.~gradients) were submitted to AWS
Batch as an array job and communication between workers of an array job
is not possible. Here, we submit a single workload (i.e.~one gradient)
as a multi-node AWS Batch job, in which case IP-based communication
between instances is enabled. Since this involves distributed memory
parallelism, we use domain decomposition based on message passing (MPI)
to solve the wave equations on multiple EC2 instances
\cite{Valli1999, MultiNode2019}. The code with the corresponding MPI
communication statements is automatically generated by the Devito
compiler. Furthermore, we use multi-threading on each individual
instance and utilize the maximum number of available cores per socket,
which is 24 for the \texttt{r5} instance and 18 for the \texttt{c5n}
instance.

\begin{figure}[!tb]
\centering
\subfloat[\label{f8a}]{\includegraphics[width=0.490\hsize]{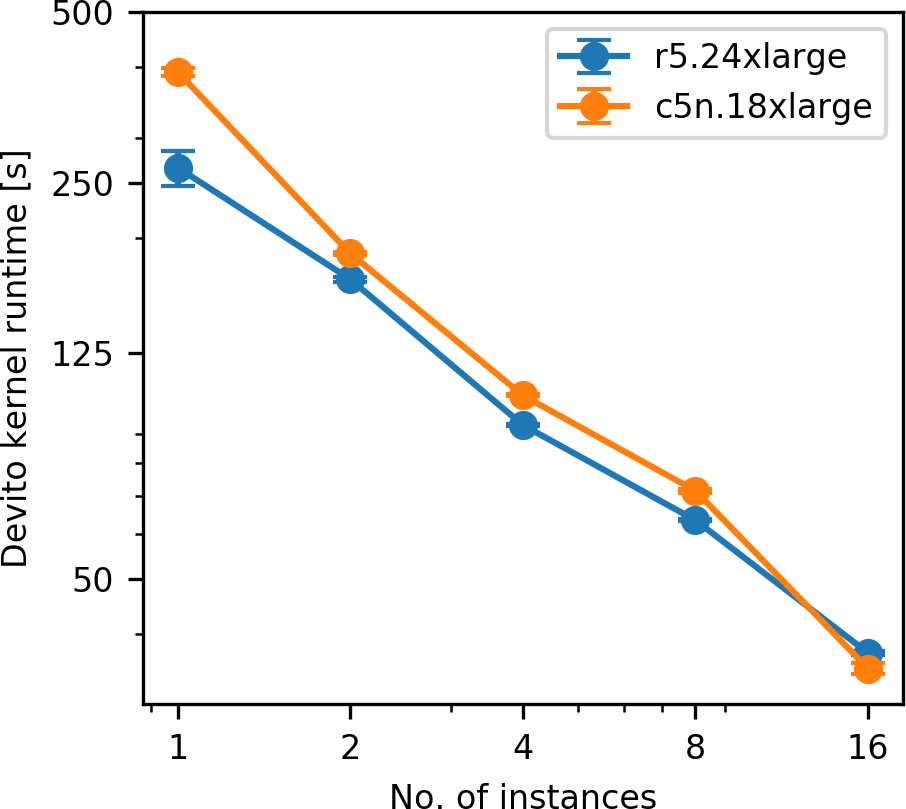}}
\hspace*{.6cm}
\subfloat[\label{f8b}]{\includegraphics[width=0.450\hsize]{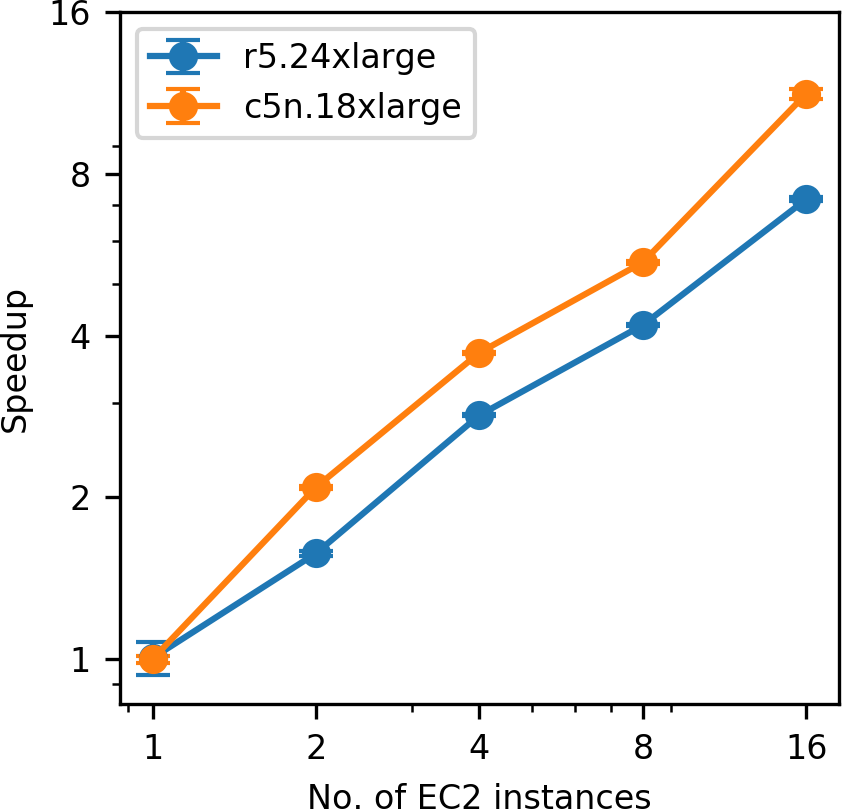}}
\\
\subfloat[\label{f8c}]{\includegraphics[width=0.469\hsize]{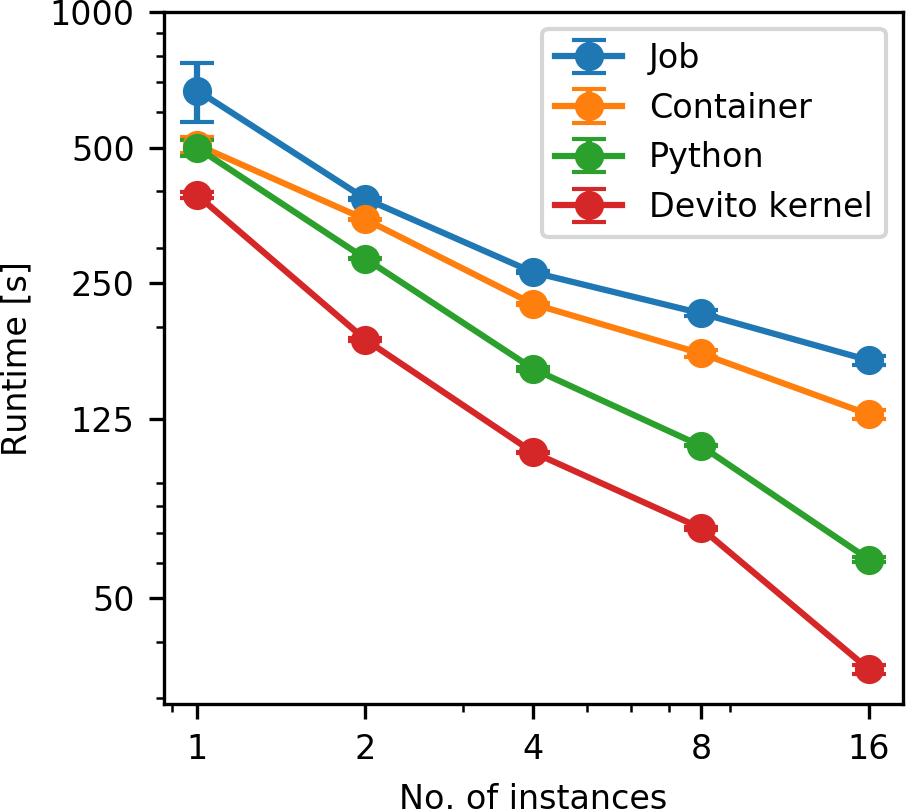}}
\hspace*{.6cm}
\subfloat[\label{f8d}]{\includegraphics[width=0.469\hsize]{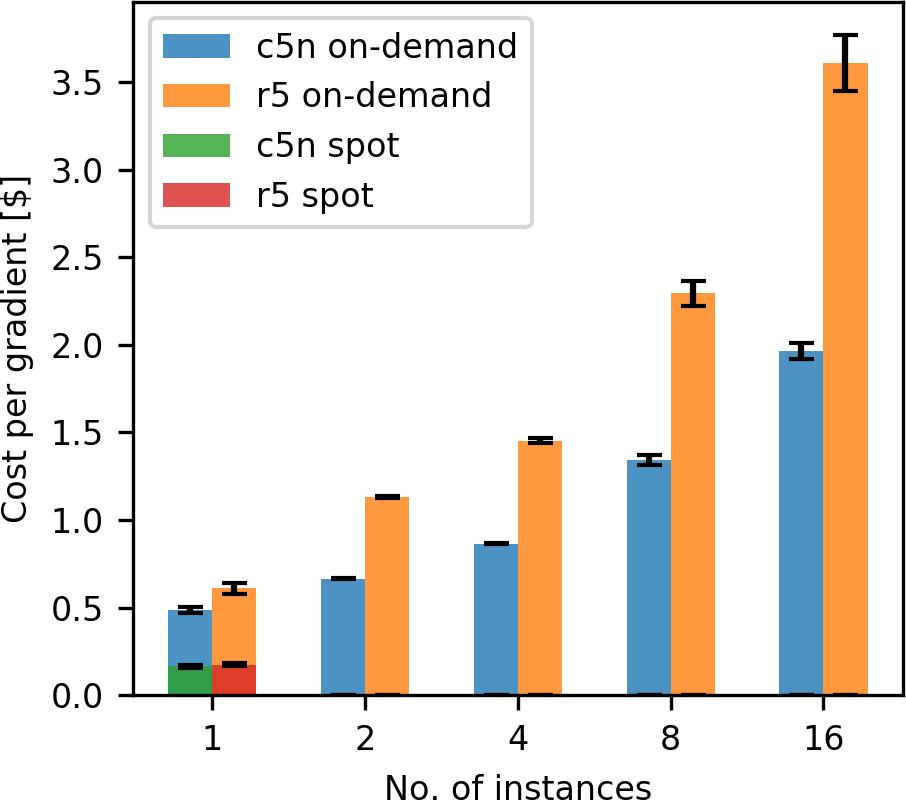}}
\caption{Strong scaling results for computing a single gradient as an
AWS Batch multi-node job for an increasing number of instances. Figures
(a) and (b) show the Devito kernel times and speedups on two different
instance types. The observed speedups are 11.3 for the \texttt{c5n}
and 7.2 for the \texttt{r5} instance. Figure (c) shows a breakdown of
the time-to-solution of each batch job into its individual components.
Figure (d) shows the EC2 cost for computing the gradients. The spot
price is only provided for the single-instance batch jobs, as spot
instances are not supported for multi-node batch jobs.}\label{f8}
\end{figure}

We compare the \texttt{r5.24xlarge} instance type from the last section
with Amazon's recently introduced \texttt{c5n} HPC instance.
Communication between AWS instances is generally based on ethernet and
the \texttt{r5} instances have up to 25 GBps networking performance.
The \texttt{c5n} instance type uses Intel Xeon Platinum 8142M processors
with up to 3.4 Ghz architecture and according to AWS provides up to
100 GBps of network bandwidth. The network is based on AWS' Nitro card
and the elastic network adapter, but AWS has not disclosed whether this
technology is based on Infiniband or Ethernet \cite{Instance2019}.
Figures~\ref{f8a} and~\ref{f8b} show the kernel runtimes and the
corresponding speedups ranging from 1 instance to 16 instances. The
\texttt{r5} instance has overall shorter runtimes than the \texttt{c5n}
instance, since the former has 24 physical cores per CPU socket, while
the \texttt{c5n} instance has 18. However, as expected, the
\texttt{c5n} instance type exhibits a better speedup than the
\texttt{r5} instance, due to the better network performance. Overall,
the observed speed up on both instances types is excellent, with the
\texttt{c5n} instance archiving a maximum speedup of 11.3 and the
\texttt{r5} instance of 7.2.

The timings given in Figure~\ref{f8a} are once again the pure kernel
times for solving the PDEs, but a breakdown of the components that make
up the total time-to-solution on the \texttt{c5n} instance is provided
in Figure~\ref{f8c}. The job runtime is defined as the interval between
the time stamp at which the batch job was created and the S3 time stamp
of the computed gradient. As in our weak scaling test, this includes the
time for AWS Batch to request and launch the EC2 instances and to start
the Docker containers, but excludes the gradient summation time, since
we are only considering the computation of a single gradient. The
container runtime is the runtime of the Docker container on the master
node and includes the time it takes AWS Batch to launch the remaining
workers and to establish an \texttt{ssh} connection between all
instances/containers. Currently, AWS Batch requires this process to be
managed by the user using a shell script that is run inside each
container. After a connection to all workers has been established, the
containers run the application as a Python program on each worker. The
Python runtime in Figure~\ref{f8c} is defined as the runtime of Python
on the main node and includes reading the seismic data from S3,
allocating memory and Devito's code generation. Our timings in
Figure~\ref{f8c} show that the overhead from requesting instances and
establishing a cluster, i.e.~the difference between the Python and
container runtime, is reasonable for a small number of instances (less
than 2 minutes), but grows significantly as the number instances is
increased to 8 and 16. Depending on the runtime of the application, the
overhead thus takes up a significant amount of the time-to-solution. In
our example, this was the case for 8 and 16 instances, but for more
compute-heavy applications that run for one or multiple hours, this
amount of overhead may still be acceptable.

Figure~\ref{f8d} shows the cost for running our scaling test as a
function of the cluster size. The cost is calculated as the instance
price (per second) times the runtime of the container on the main node
times the number of instances. The cost per gradient grows significantly
with the number of instances, as the overhead from establishing an
\texttt{ssh} connection to all workers increases with the cluster size.
The communication overhead during domain decomposition adds an
additional layer of overhead that further increases the cost for an
increasing number of instances. This is an important consideration for
HPC in the cloud, as the shortest time-to-solution does not necessarily
correspond to the cheapest approach. Another important aspect is that
AWS Batch multi-node jobs do not support spot instances
\cite{MultiNode2019}. Spot instances allow users to access unused EC2
capacities at significantly lower price than at the on-demand price, but
AWS can terminate spot instances at any time with a two minute warning,
e.g.~if the demand for that instance type increases \cite{Spot2019}.
Spot instances are typically in the range of 2 to 3 times cheaper than
the corresponding on-demand price, but AWS Batch multi-node jobs are,
for the time being, only supported with on-demand instances.

\begin{figure}[!tb]
\centering
\subfloat[\label{f9a}]{\includegraphics[width=0.470\hsize]{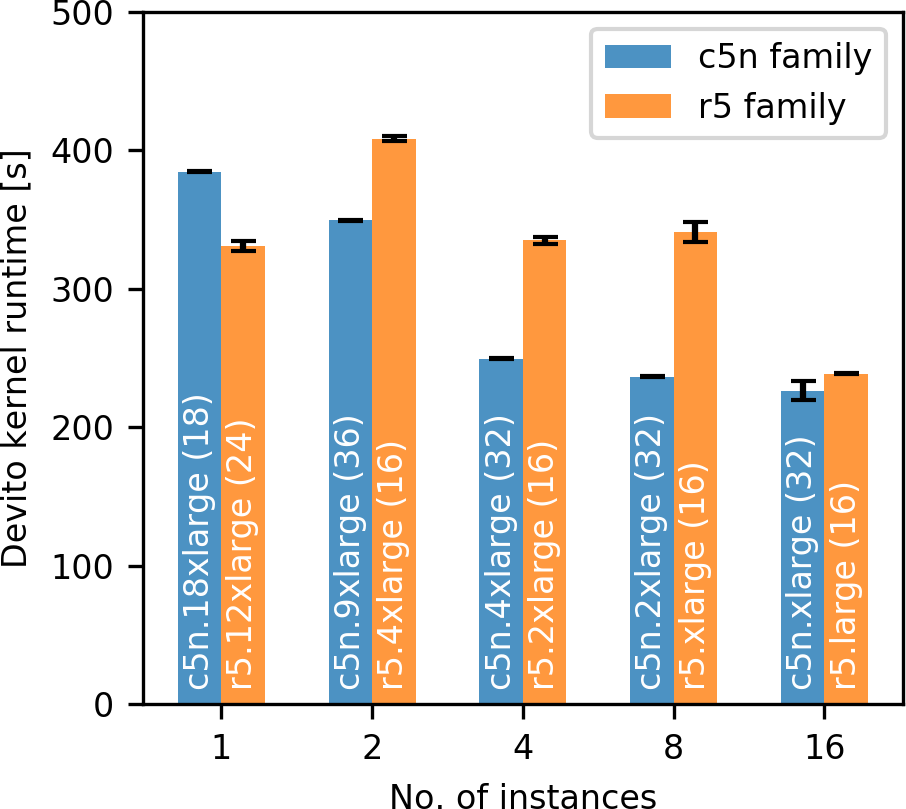}}
\hspace*{.6cm}
\subfloat[\label{f9b}]{\includegraphics[width=0.470\hsize]{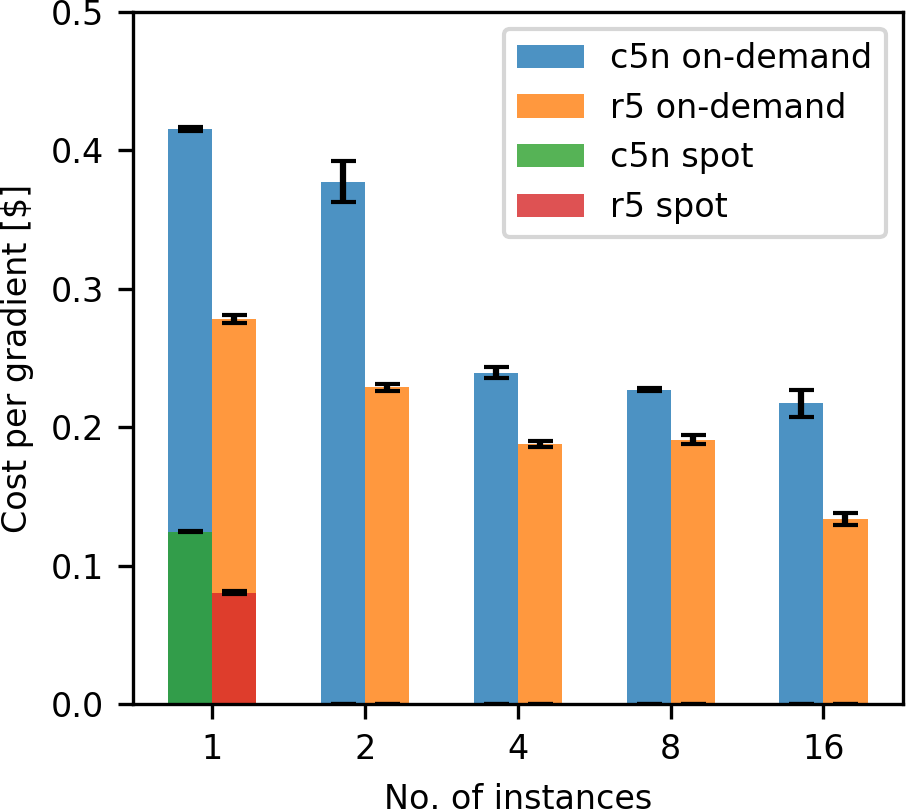}}
\caption{Devito kernel runtimes for computing a single gradient as an
AWS Batch job for an increasing number of instances. In comparison to
the previous example, we use the smallest possible instance type for
each job, as specified in each bar. We use the maximum number of
available cores on every instance type and the total number of cores
across all instances is given in each bar. Figure (b) shows the
corresponding cost for computing the gradients. Choosing the instance
size such the total memory is approximately constant, avoids that the
cost increases as a larger number of instances are used per gradient.
However, single instances using spot prices ultimately remain the
cheapest option.}\label{f9}
\end{figure}

The scaling and cost analysis in Figures~\ref{f8a} --~\ref{f8d} was
carried out on the largest instances of the respective instance types
(\texttt{r5.24xlarge} and \texttt{c5n.18xlarge}) to guarantee exclusive
access to the compute nodes and network bandwidth. Increasing the number
of instances per run therefore not only increases the total number of
available cores, but also the amount of memory. However, for computing a
single gradient, the required amount of memory is fixed, so increasing
the number of instances reduces the required amount of memory per
instance, as wavefields are distributed among more workers. In practice,
it therefore makes sense to chose the instance type based on the
required amount of memory per worker, as memory is generally more
expensive than compute. In our specific case, computing the gradient
requires 170 GB of memory, which requires either a single
\texttt{r5.12xlarge} instance, two \texttt{r5.4xlarge}, four
\texttt{r5.2xlarge}, eight \texttt{r5.xlarge} or sixteen
\texttt{r5.large} instances. However, these instances not only differ in
the amount of memory, but also in the number of CPU cores. We repeat our
previous scaling test, but rather than using the same instance type in
all runs, we choose the instance type based on the required amount of
memory. Furthermore, for every instance type, we utilize the maximum
amount of available cores using multi-threading with OpenMP. The kernel
runtimes for an increasing number of instances is shown in
Figure~\ref{f9a}. In each bar, we indicate which instance type was used,
as well as the total number of cores across all instances. The
corresponding costs for computing each gradient is shown in
Figure~\ref{f9a}. Compared to the previous example, we observe that
using 16 small on-demand instances leads to a lower cost than using a
single more expensive large instance, but that using a single instance
ultimately remains the most cost-effective way of computing a gradient,
due to the possibility to utilize spot instances.

\begin{table}
\renewcommand{\arraystretch}{1.3}
\caption{Comparison of parallelization strategies on a single EC2
instance in the context of AWS Batch. The timings are the Devito kernel
times for computing a single gradient of the BP model using AWS Batch.
The program runs as a single docker container on an individual EC2
instance, using either multi-threading (OpenMP) or a hybrid approach of
multithreading and domain-decomposition (OpenMP + MPI).}\label{t1}
\centering
\begin{tabular}{l l l l l}
\hline
Grid & CPU (cores) & Parallelization & Runtime
{[}s{]}\tabularnewline
\hline
$1,911 \times 5,394$ & $1$ $(24)$ & OMP &
$190.17 \pm 7.12$\\
$1,911 \times 10,789$ & $1$ $(24)$ & OMP &
$378.94 \pm 13.57$\\
$1,911 \times 10,789$ & $2$ $(48)$ & OMP &
$315.92 \pm 16.50$\\
$1,911 \times 10,789$ & $2$ $(48)$ & OMP + MPI &
$249.13 \pm 5.22$\\
\hline
\end{tabular}

\end{table}

In terms of cost, our scaling examples underline the importance of
choosing the EC2 instances for the AWS Batch jobs based on the total
amount of required memory, rather than based on the amount of CPU cores.
Scaling horizontally by using an increasingly large number of instances
expectedly leads to a faster time-to-solution, but results in a
significant increase of cost as well (Figure~\ref{f8d}). As shown in
Figure~\ref{f9b}, this increase in cost can be avoided to some extent by
choosing the instance size such that the total amount of memory stays
approximately constant, but ultimately the restriction of not supporting
spot instances, makes multi-node batch jobs not attractive in scenarios
where single instances provide sufficient memory to run a given
application. In practice, it makes therefore sense to use single
node/instance batch jobs and to utilize the full number of available
cores on each instance. The largest EC2 instances of each type (e.g.
\texttt{r5.24xlarge}, \texttt{c5n.18xlarge}) have two CPU sockets with
shared memory, making it possible to run parallel programs using either
pure multi-threading or a hybrid MPI-OpenMP approach. In the latter
case, programs still run as a single Docker container, but within each
container use MPI for communication between CPU sockets, while OpenMP is
used for multithreading on each CPU. For our example, we found that
computing a single gradient of the BP model with the hybrid MPI-OpenMP
approach leads to a 20 \% speedup over the pure OpenMP program
(Table~\ref{t1}), which correspondingly leads to
20 \% cost savings as well.

\subsection{Cost comparison}\label{cost-comparison}

One of the most important considerations of high performance computing
in the cloud is the aspect of cost. As users are billed for running EC2
instances by the second, it is important to use instances efficiently
and to avoid idle resources. This is oftentimes challenging when running
jobs on a conventional cluster. In our specific application, gradients
for different seismic source locations are computed by a pool of
parallel workers, but as discussed earlier, computations do not
necessarily complete at the same time. On a conventional cluster,
programs with a MapReduce structure, such as parallel gradient
computations, are implemented based on a client-server model, in which
the workers (i.e.~the clients) compute the gradients in parallel, while
the master (the server) collects and sums the results. This means that the
process has to wait until all gradients $\mathbf{g}_i$ have been
computed, before the gradient can be summed and used to update the
image. This inevitably causes workers that finish their computations
earlier than others to the sit idle. This is problematic when using a
cluster of EC2 instances, where the number of instances are fixed, as
users have to pay for idle resources. In contrast, the event-driven
approach based on Lambda functions and AWS Batch automatically
terminates EC2 instances of workers that have completed their gradient
calculation, thus preventing resources from sitting idle. This
fundamentally shifts the responsibility of requesting and managing the
underlying EC2 instances from the user to the cloud environment and
leads to significant cost savings as demonstrated in the following
example.

\begin{figure}[!tb]
\centering
\subfloat[\label{f10a}]{\includegraphics[width=0.444\hsize]{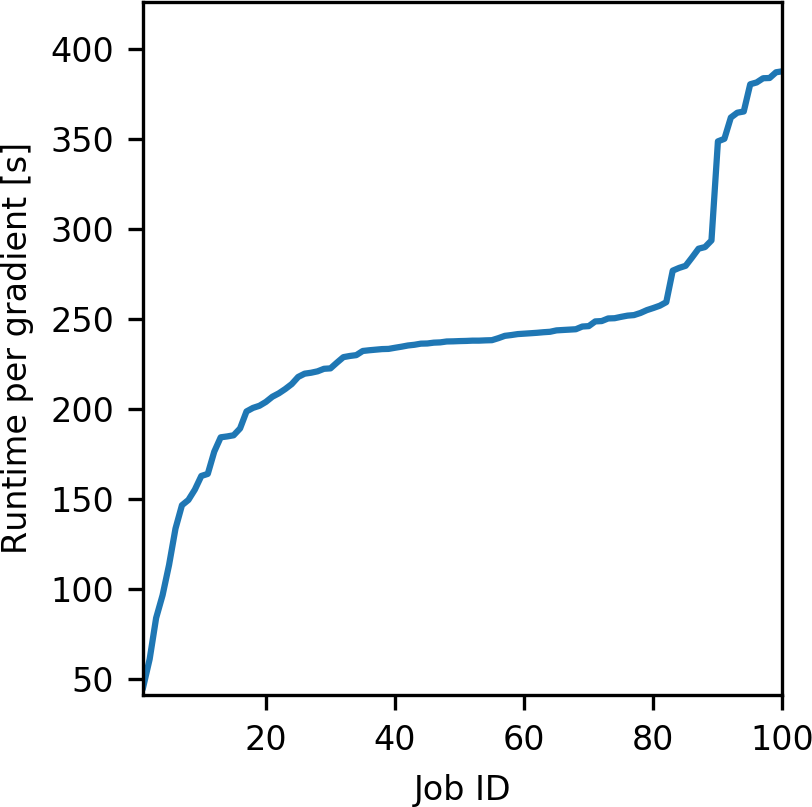}}
\hspace*{.6cm}
\subfloat[\label{f10b}]{\includegraphics[width=0.496\hsize]{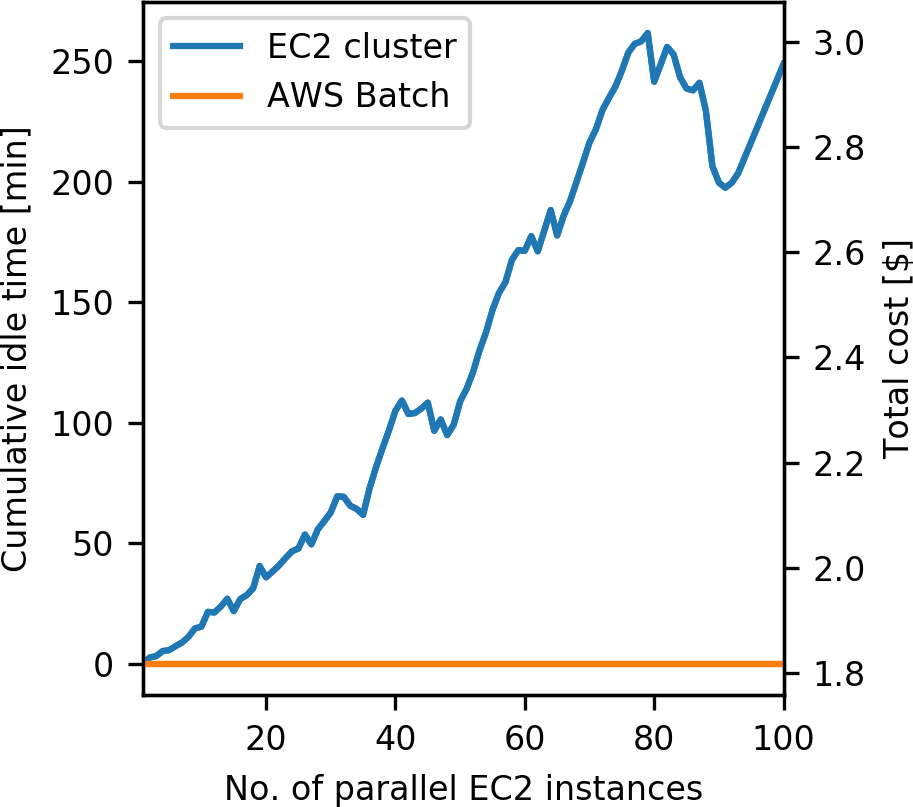}}
\\
\subfloat[\label{f10c}]{\includegraphics[width=0.470\hsize]{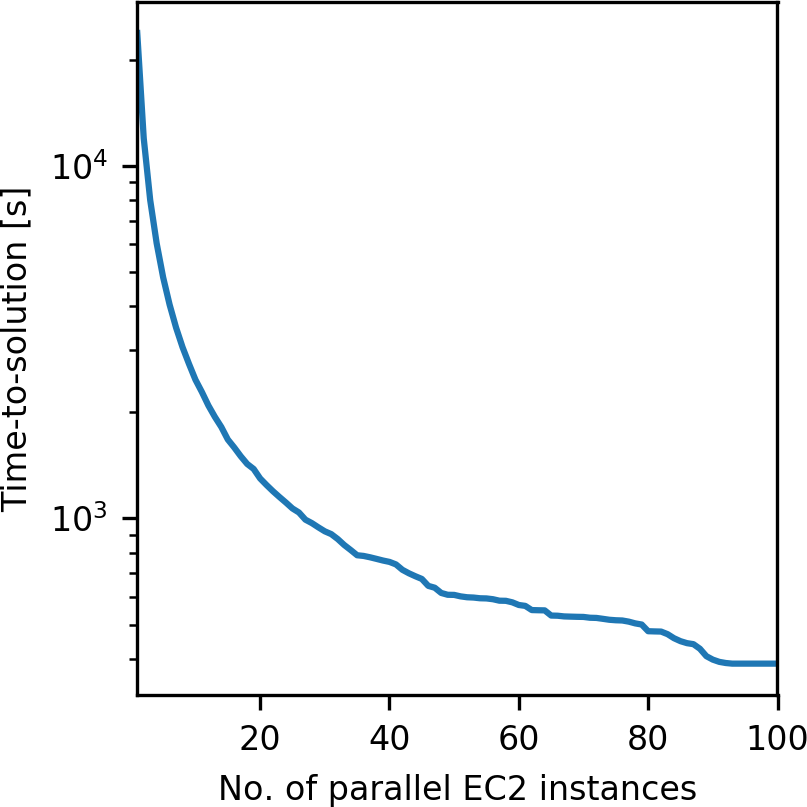}}
\caption{(a) Sorted container runtimes of an AWS Batch job in which we
compute the gradient of the BP model for a batch size of 100. Figure (b)
shows the cumulative idle time for computing this workload as a function
of the number of parallel workers on either a fixed cluster of EC2
instances or using AWS Batch. The right-hand y-axis shows the
corresponding cost, which is proportional to the idle time. In the
optimal case, i.e.~no instances every sit idle, the cost for computing a
gradient of batch size 100 is 1.8\$. Figure (c) shows the
time-to-solution as a function of the number of parallel instances,
which is the same on an EC2 cluster and for AWS Batch, if we ignore the
startup time of the AWS Batch workers or of the corresponding EC2
cluster.}\label{f10}
\end{figure}

We illustrate the difference between the event-driven approach and using
a fixed cluster of EC2 instances by means of a specific example. We
consider our previous example of the BP synthetic model and assume that
we want to compute the gradient for a batch size of 100. As in our weak
scaling experiment, we restrict the modeling domain to the subset of the
model that includes the respective seismic source location, as well as
the seismic receivers that record the data. Towards the edge of the
model, the modeling domain is smaller, as some receivers lie outside the
modeling domain and are therefore omitted. We compute the gradient
$\mathbf{g}_i$ for 100 random source locations and record the runtimes
(Figure~\ref{f10a}). We note that most gradients take around 250 seconds
to compute, but that the runtimes vary due to different domain sizes and
varying EC2 capacity (similar to the timings in Figure~\ref{f5c}). We
now model the idle times for computing these gradients on a cluster of
EC2 instances as a function of the the number of parallel instances,
ranging from 1 instance (fully serial) to 100 instances (fully
parallel). For a cluster consisting of a single instance, the cumulative
idle time is naturally zero, as the full workload is executed in serial
by a single instance. For more than one instance, we model the amount of
time that each instance is utilized, assuming that the workloads are
assigned dynamically to the available instances. The cumulative idle
time $t_\text{idle}$ is then given as the sum of the differences between
the runtime of each individual instance $t_i$ and the instance with the
longest runtime:
\begin{equation}
t_\text{idle} = \sum_{i=1}^{n_{\text{EC2}}} (\max\{t_i\} - t_i),
\label{idletime}
\end{equation}
 The cumulative idle time as a function of the cluster size
$n_\text{EC2}$ is plotted in Figure~\ref{f10b}. We note that the
cumulative idle time generally increases with the cluster size, as a
larger number of instances sit idle while waiting for the final gradient
to be computed. On a cluster with 100 instances each gradient is
computed by a separate instance, but all workers have to wait until the
last worker finishes its computation (after approximately 387 seconds).
In this case, the varying time-to-solutions of the individual gradients
leads to a cumulative idle time of 248 minutes. Compared to the
cumulative computation time of all gradients, which is 397 minutes, this
introduces an overhead of more than 60 percent, if the gradients are
computed on a cluster with 100 instances. The cumulative idle time is
directly proportional to the cost for computing the 100 gradients, which
is plotted on the right axis of Figure~\ref{f10b}. With AWS Batch, the
cumulative idle time for computing the 100 gradients is zero, regardless
of the number of parallel instances that AWS Batch has access to. Any
EC2 instance that is not utilized anymore is automatically shut down by
AWS Batch, so no additional cost other than the pure computation time of
the gradients is invoked \cite{BatchShutDown2019}.

In practice, it is to be expected that the cost savings of AWS Batch are
even greater, as we are not taking the time into account that it takes
to start an EC2 cluster of a specified number of instances. In our weak
scaling experiments (Figure~\ref{f5a}), we found that spinning up a
large number of EC2 instances does not happen instantaneously but over a
period of several minutes, so starting a cluster of EC2 instances
inevitably causes some instances to sit idle while the remaining
instances are started. This was also observed for our multi-node AWS
Batch job experiment (Figure~\ref{f8c}), but in this case the cluster
size per gradient is considerably smaller than the size of a single
large cluster for computing all gradients. Single-node AWS Batch jobs do
not suffer from the variable startup time, as workers that are launched
earlier than others instantaneously start their computations, without
having to wait for the other instances.

While computing the 100 gradients on an EC2 cluster with a small number
of instances results in little cumulative idle time, it increases the
overall time-to-solution, as a larger number of gradients have to be
sequentially computed on each instance~\eqref{f10c}. With AWS Batch this
trade-off does not exist, as the cumulative idle time, and therefore the
cost for computing a fixed workload, does not depend on the number of
instances. However, it is to be expected that in practice the
time-to-solution is somewhat larger for AWS Batch than for a fixed
cluster of EC2 instances, as AWS Batch needs to request and launch EC2
instances for every new gradient computation. In our weak scaling
experiments in which we requested up to 128 instances, we found that the
corresponding overhead lies in the range of 3 to 10 minutes (per
iteration), but it is to be expected that the overhead further grows for
an even larger number of instances. However, no additional cost is
introduced while AWS Batch waits for the EC2 instances to start.

\subsection{Cost saving strategies for AWS
Batch}\label{cost-saving-strategies-for-aws-batch}

Using AWS Batch for computing the gradients of a seismic imaging
workflow limits the runtime that each EC2 instance is active to the
amount of time it takes to compute a single gradient (of one source
location). Running a seismic imaging workflow on a conventional cluster
of EC2 instances, requires instances to stay up during the entire
execution time of the program, i.e.~for all iterations of the seismic
imaging optimization algorithm. The limitation of instance runtimes to
the duration of a single gradient computations with AWS Batch is
beneficial for the usage of spot instances, as it reduces the chance
that a specific instance is shut down within the duration it is used. As
demonstrated in our earlier examples (Figures~\ref{f8d} and~\ref{f9b}),
spot instances can significantly reduce the cost of running EC2
instances, oftentimes by a factor of 2-4 in comparison to on-demand
instances (Table~\ref{t2}).

\begin{table}
\renewcommand{\arraystretch}{1.3}
\caption{AWS on-demand and spot prices of a selection of EC2 instance
types that were used in the previous experiments. Prices are provided
for the US East (North Virginia) region, in which all experiments were
carried out (07/11/2019).}\label{t2}
\centering
\begin{tabular}{l l l l}
\hline
Instance & On-demand (\$/hour) & Spot (\$/hour) & Ratio \\
\hline
m4.4xlarge & $0.800$ & $0.2821$ & $2.84$ \\
r5.24xlarge & $6.048$ & $1.7103$ & $3.54$ \\
c5n.18xlarge & $3.888$ & $1.1659$ & $3.33$ \\
\hline
\end{tabular}

\end{table}

\begin{figure}[!tb]
\centering
\subfloat[\label{f11a}]{\includegraphics[width=0.460\hsize]{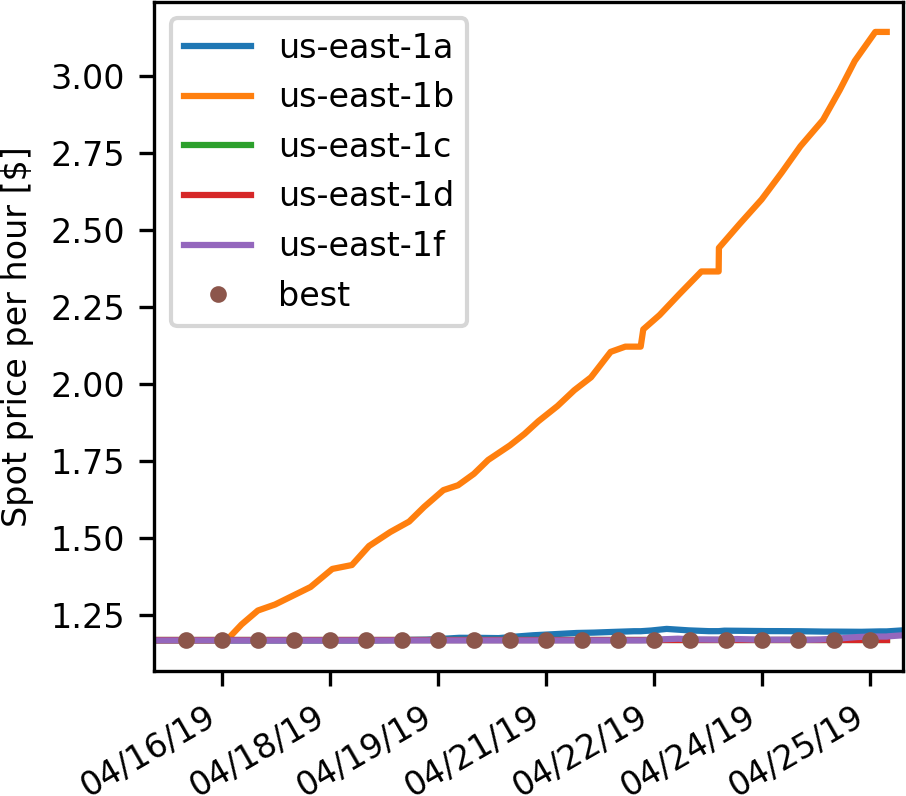}}
\hspace*{.6cm}
\subfloat[\label{f11b}]{\includegraphics[width=0.460\hsize]{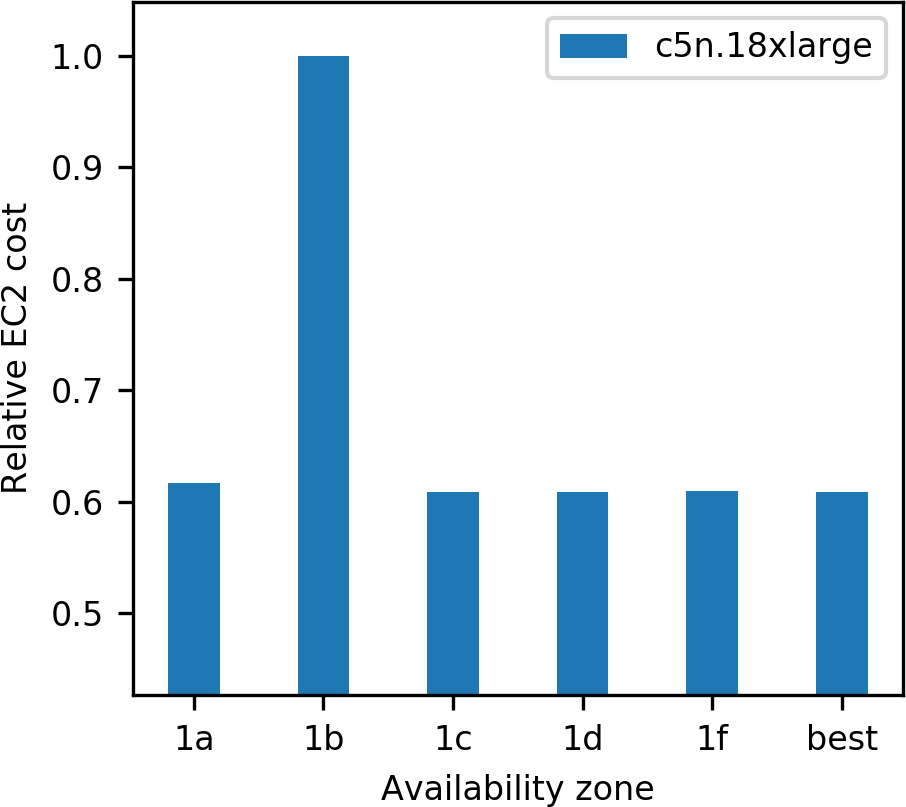}}
\\
\subfloat[\label{f11c}]{\includegraphics[width=0.460\hsize]{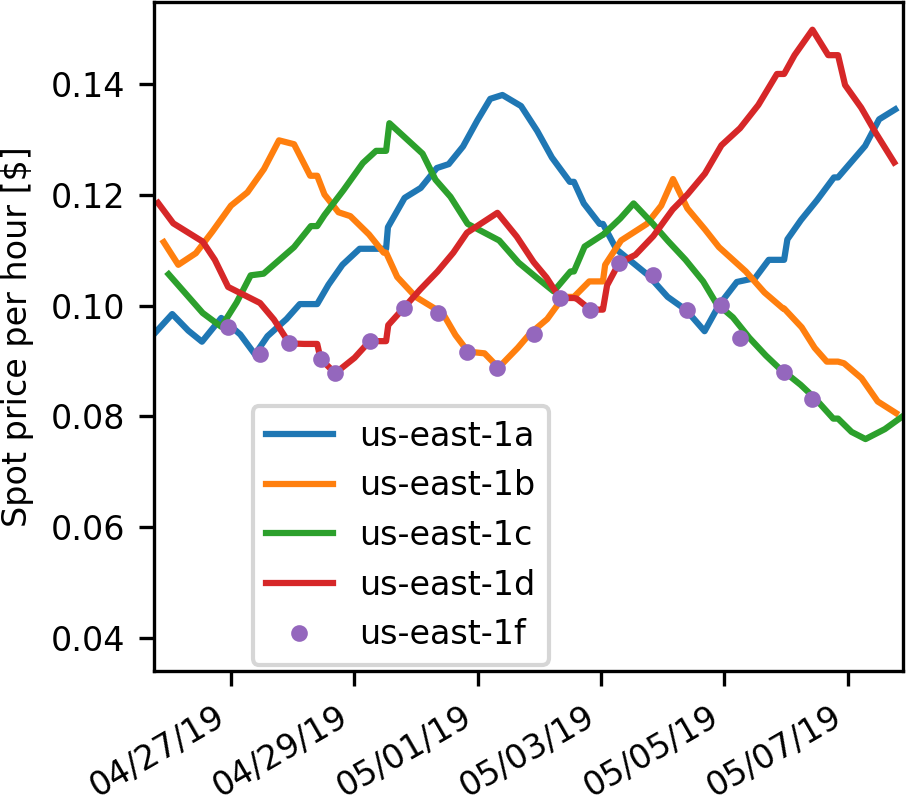}}
\hspace*{.6cm}
\subfloat[\label{f11d}]{\includegraphics[width=0.460\hsize]{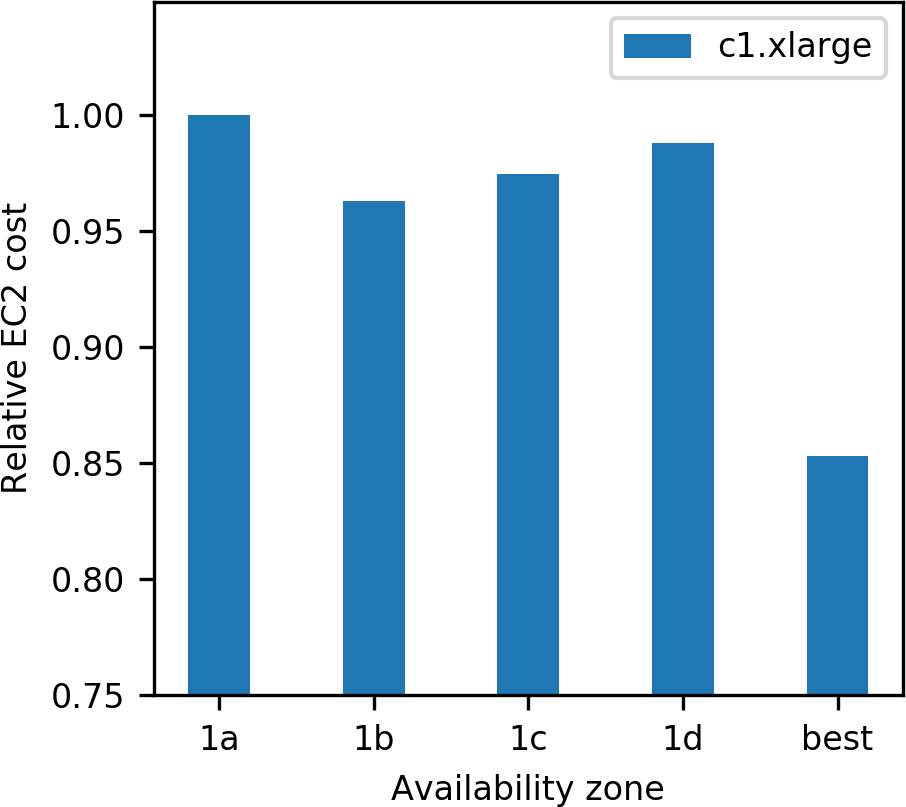}}
\caption{(a) Historical spot price of the \texttt{c5n.18xlarge} instance
in different zones of the US East region over a 10 ten day period in
April 2019. Figure (b) shows the relative cost for running an iterative
seismic imaging algorithm over this time period in the respective zones.
The right-most bar indicates the price for running the application with
our event-driven workflow, in which the cheapest zone is automatically
chosen at the start of each iteration (indicated as dots in Figure a).
Figures (c) and (d) are the same plots for the \texttt{c1.xlarge}
instance during a different time window.}\label{f11}
\end{figure}

In contrast to on-demand instances, the price of spot instances is not
fixed and depends on the current demand and availability of the instance
type that is being requested. As such, prices for spot instances can
vary significantly between the different zones of a specific region
(Figure~\ref{f11a}). If spot instances are used for a cluster and are
fixed for the entire duration of the program execution, users are
exposed to variations of the spot price during that time period. This
effects is usually negligible for programs that run in a matter of
hours, as spot prices typically do not vary substantially over short
periods of time. However, large-scale 3D seismic imaging problems
potentially run over the course of multiple days, in which case varying
spot prices can have significant influence on the cost.

We consider a hypothetical example in which we assume that we run a
large-scale imaging example for 20 iterations of an optimization
algorithm, where each iteration takes 12 hours to compute, which leads
to a total runtime of 10 days. Figures~\ref{f11a} shows the historical
spot price of the \texttt{c5n.18xlarge} instance over a 10 day time
period in April 2019 and Figure~\ref{f11b} shows the normalized cost of
running the example in each specific zone. We note that the spot price
of this instance type is the same across all zones at the start of the
program, but that the spot price in the \texttt{us-east-1b} zone starts
increasing significantly after a few hours. In this case, running the
example on a fixed cluster of instances results in cost differences of
40 percent, depending on which zone was chosen at the start of the
program. Our event-driven workflow with AWS Batch, allows that the
cheapest available zone is automatically chosen at the start of every
iteration, which ensures that zones that exhibit a sudden increase of
the spot price, such as zone \texttt{us-east-1b}, are avoided in the
subsequent iteration.

Another example for a different time interval and the \texttt{c1.xlarge}
instance type is shown in Figure~\ref{f11c} and the relative cost of
running the example in the respective zone is plotted in
Figure~\ref{f11d}. The right-most bar shows the cost if the example was
run with AWS Batch and the event-driven workflow had chosen the cheapest
available zone at each iteration. In this case, switching zones between
iterations leads to cost savings of 15 percent in comparison to running
the example in the \texttt{us-east-1a} zone, which is the cheapest zone
at the start of the program. For our cost estimates (Figure~\ref{f11b}
and~\ref{f11d}), we assume that the spot price is not affected by our
own application, i.e.~by our own request of a potentially large number
of spot instances, but in practice this issue needs to be taken into
account as well. Overall, the two examples shown here are obviously
extreme cases of price variations across zones and in practice the spot
price is oftentimes reasonably stable. However, spot price fluctuations
are nevertheless unpredictable and the event-driven approach with AWS
Batch allows to minimize exposure to price variations.

\begin{figure}[!tb]
\centering
\subfloat[\label{f12a}]{\includegraphics[width=0.460\hsize]{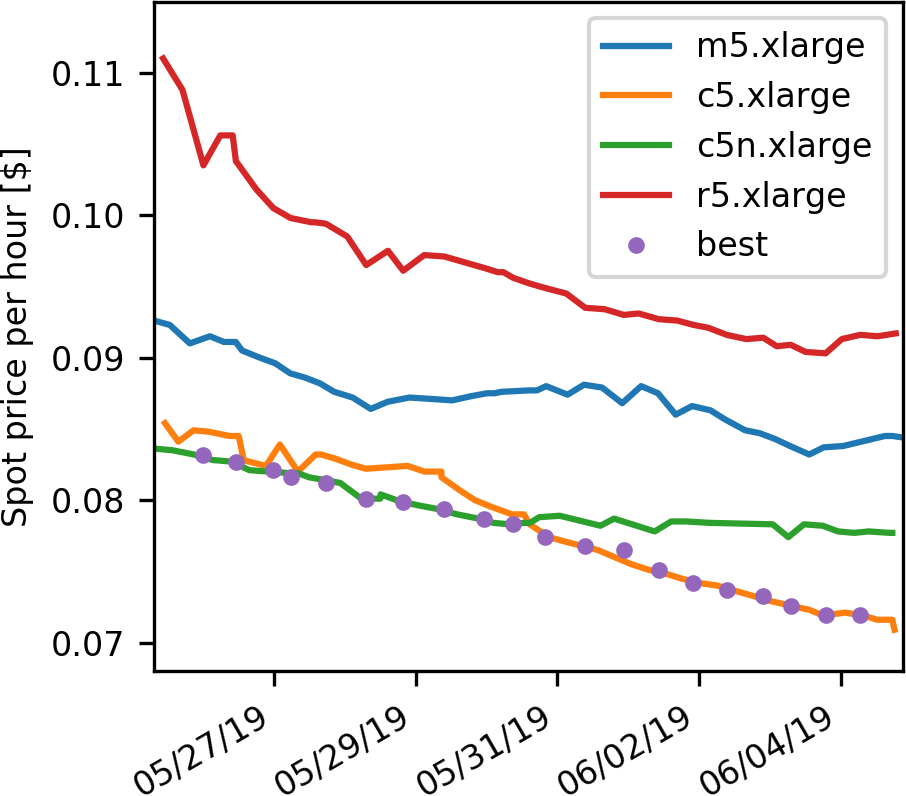}}
\hspace*{.6cm}
\subfloat[\label{f12b}]{\includegraphics[width=0.460\hsize]{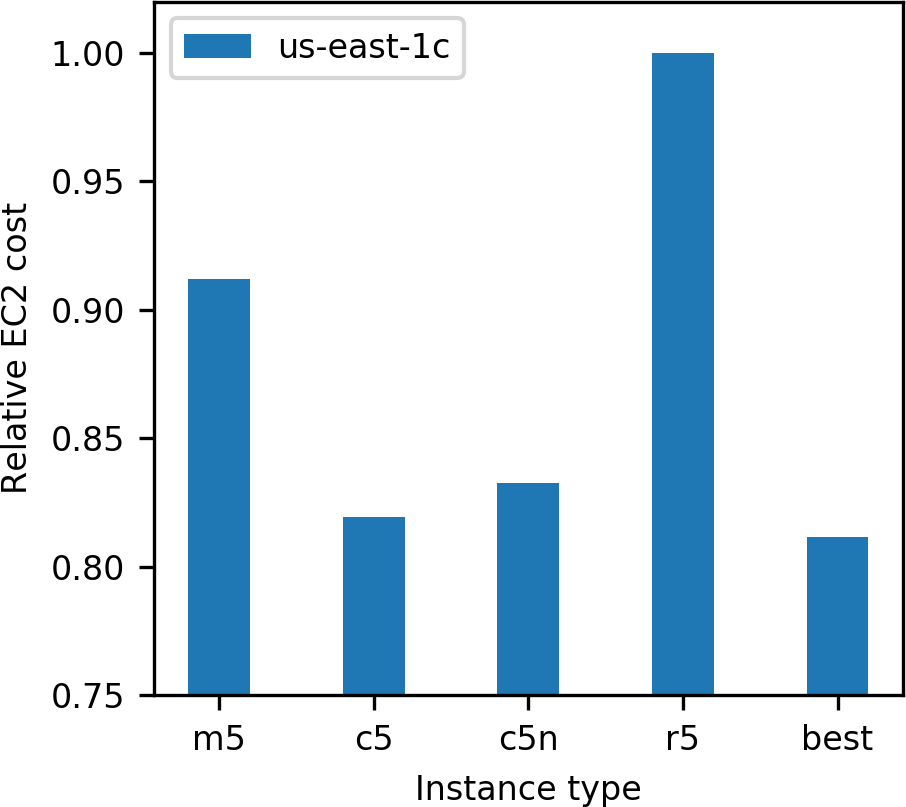}}
\caption{(a) Historical spot prices for a variety of \texttt{4xlarge}
instances over a 10 day period in April 2019. All shown instances have
$128$ GB of memory, but vary in their number of CPU cores and
architectures. Figure (b) shows the relative cost of running an
iterative seismic imaging application over this time period in the
respective zone and for the case, in which the cheapest available
instance is chosen at the beginning of each iteration.}\label{f12}
\end{figure}

Apart from choosing spot instances in different zones, it is also
possible to vary the instance type in that is used for the computations.
For example, in Figure~\ref{f12a}, we plot the historical spot price of
the \texttt{xlarge} instance for various instance types (\texttt{m5},
\texttt{c5}, \texttt{c5n} and \texttt{r5}). All instances have 4 virtual
CPUs, but vary in the amount of memory and their respective CPU
architecture. Spot instances are not priced proportionally to their
hardware (memory, cores, architecture), but based on the current demand.
Therefore, it is oftentimes beneficial to compare different instance
types and choose the currently cheapest type from a pool of possible
instances. As before, we compare the relative cost for running the 10
day example on a cluster of EC2 instances, in which case the instance
type is fixed for the duration of the program, against the dynamic
approach with AWS Batch. Again, the event-driven approach allows to
minimize exposure to price changes over the duration of the example, by
choosing the cheapest available instance type at the beginning of each
iteration.

\subsection{Resilience}\label{resilience_section}

In the final experiment of our performance analysis, we analyze the
resilience of our workflow and draw a comparison to running an MPI
program on a conventional cluster of EC2 instances. Resilience is an
important factor in high performance computing, especially for
applications like seismic imaging, whose runtime can range from several
hours to multiple days. In the cloud, the mean-time-between failures is
typically much shorter than on comparable HPC systems
\cite{jackson2010}, making resilience potentially a prohibiting factor.
Furthermore, using spot instances further increases the exposure to
instance shut downs, as spot instances can be terminated at any point in
time with a two minute warning.

Seismic imaging codes that run on conventional HPC clusters typically
use MPI to parallelize the sum of the source indices. MPI based
applications exhibit a well known shortcoming of having a relatively low
fault tolerance, as hardware failures lead to the termination of a
running program. Currently, the only noteworthy approach of fault
tolerance for MPI programs is the User Level Fault Mitigation (ULFM)
Standard \cite{bland2013}. ULFM enables an MPI program to continue
running after a node/instance failure, using the remaining available
resources to finish the program execution. Using AWS Batch to compute
the gradients $\mathbf{g}_i$ for a given batch size, provides a natural
fault tolerance, as each gradient is computed by a separate container,
so the crash of one instance does not affect the execution of the code
on the remaining workers. Furthermore, AWS Batch provides the
possibility to automatically restart EC2 instances that have crashed. In
contrast to ULFM, this allows the completion of programs with the
initial number of nodes or EC2 instances, rather than with a reduced
number.

We illustrate the effect of instance restarts by means of our previous
example with the BP model (Figure~\ref{f4}). Once again, we compute the
gradient of the LS-RTM objective function for a batch size of 100 and
record the runtimes without any instance/node failures. We compute the
gradients using two different computational strategies for
backpropagation. In the first approach, we compute the gradient on
instances with a sufficient amount of memory to store the state
variables in memory, which leads to an average runtime per gradient of 5
minutes (using 62 GB of memory). In the second approach, we compute the
gradients on instances with less memory and use optimal checkpointing
\cite{Griewank2000}, in which case we store only a small subset of
state variables and recompute the remaining states during
backpropagation. This increases the average runtime per gradient to 45
minutes, but also reduces the required amount of memory for this example
from 62 GB to 5 GB.

\begin{figure}[!tb]
\centering
\subfloat[\label{f13a}]{\includegraphics[width=0.470\hsize]{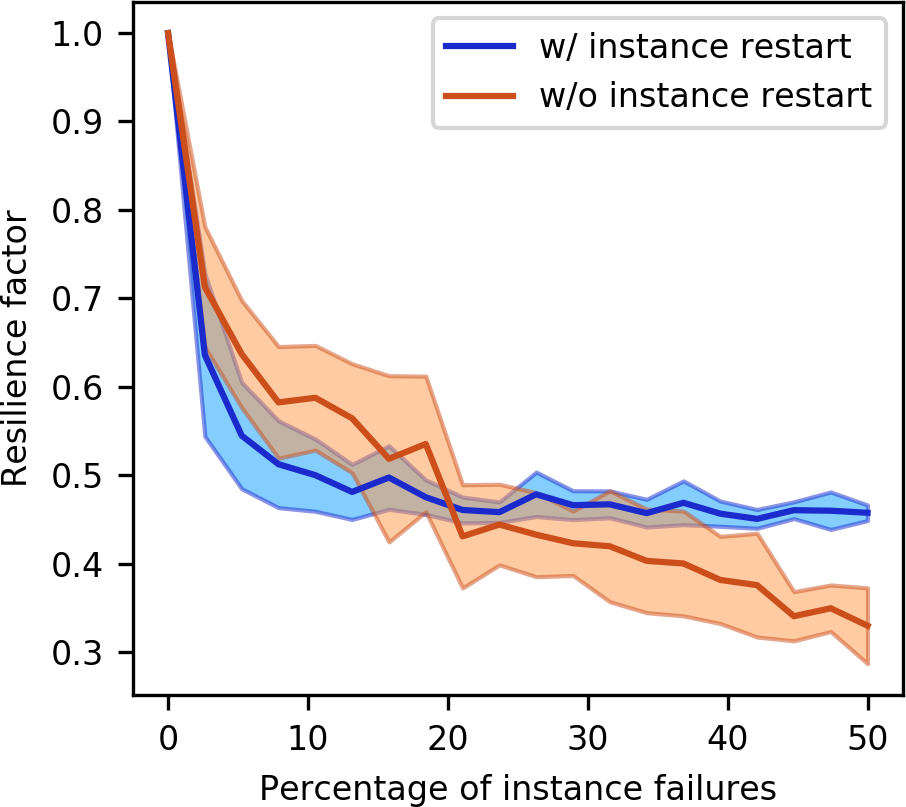}}
\hspace*{.6cm}
\subfloat[\label{f13b}]{\includegraphics[width=0.470\hsize]{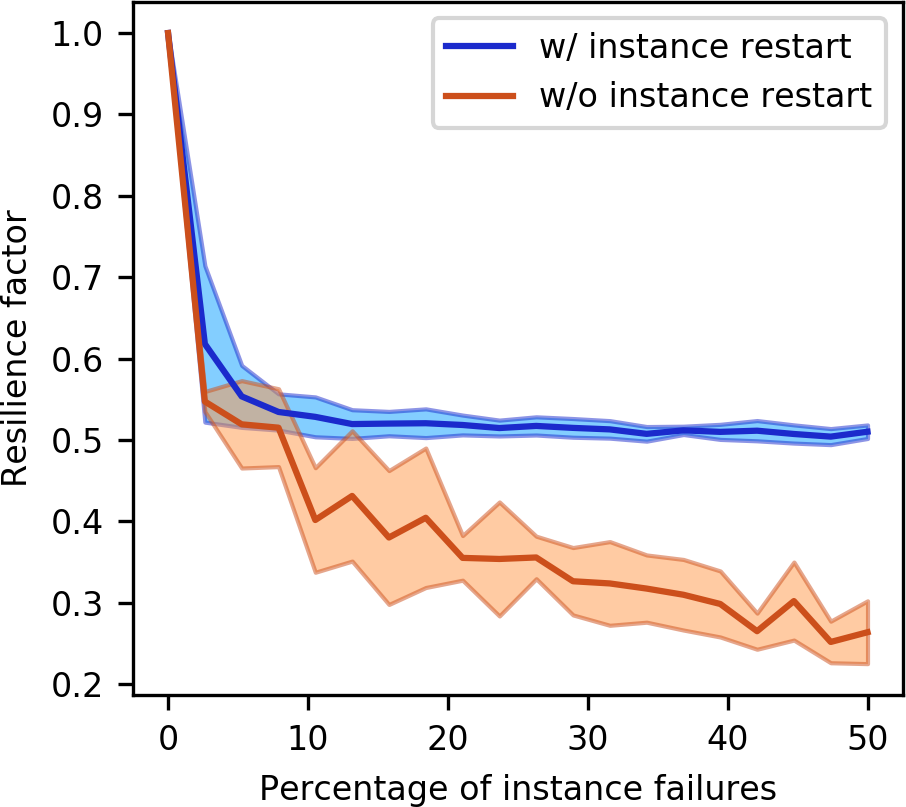}}
\caption{Comparison of the resilience factor (RF) for an increasing
percentage of instance failures with and without instance restarts. The
RF provides the ratio between the original runtime for computing the
gradient of the BP model for a batch size of 100 and the runtime in the
presence of instance failures. Figure (a) is the RF for an application
that runs for 5 minutes without failures, while figure (b) is based on
an example whose original time-to-solution is 45 minutes.}\label{f13}
\end{figure}

We then model the time that it takes to compute the 100 gradients for
an increasing number of instance failures with and without restarts. We
assume that the gradients are computed fully in parallel, i.e.~on 100
parallel instances and invoke an increasing number of instance failures
at randomly chosen times during the execution of program. Without
instance restarts, we assign the workload of the failed instances to the
workers of the remaining instances and model how long it takes complete
the computation of the 100 gradients. With restarts, we add a two
minute penalty to the failed workers and then restart the computation on
the same instance. The two minute penalty represents the average amount
of time it takes AWS to restart a terminated EC2 instance and was
determined experimentally by manually shutting down EC2 instances of an
AWS Batch job and recording the time it takes to restart the container
on a new instance.

Figure~\ref{f13} shows the ratio of the time-to-solution for computing
the 100 gradients without events (i.e.~without failures) to the
modeled time-to-solution with events. This ratio is known as the
resilience factor \cite{hukerikar2017} and provides a metric of how
instance failures affect the time-to-solution and therefore the cost of
running a given application in the cloud:
\begin{equation}
r = \frac{\text{time-to-solution }_\text{ event-free}}{\text{time-to-solution }_\text{ event}}
\label{resilience}
\end{equation}
 Ideally, we aim for this factor being as close to 1 as possible,
meaning that instance failures do not significantly increase the
time-to-solution. Figures~\ref{f13a} and~\ref{f13b} show the resilience
factors with and without restarts for the two different backpropagation
strategies, which represent programs of different runtimes. The
resilience factor is plotted as a function of the percentage of instance
failures and is the average of 10 realizations, with the standard
deviation being depicted by the shaded colors. The plots show that the
largest benefit from being able to restart instances with AWS Batch is
achieved for long running applications (Figure~\ref{f13b}). The
resilience factor with instance restarts approaches a value of 0.5,
since in the worst case, the time-to-solution is doubled if an instance
fails shortly before completing its gradient computation. Without being
able to restart instances, as would be the case for MPI programs with
ULFM, the gradient computations need to be completed by the remaining
workers, so the resilience factor continuously decreases as the failure
percentage increases. For short running applications
(Figure~\ref{f13a}), the overhead of restarting instances diminishes the
advantage of instance restarts, unless a significant percentage of
instances fail, which, however, is unlikely for programs that run in a
matter of minutes. On the other hand, long running programs or
applications with a large number of workers are much more likely to
encounter instance shut downs and our experiment shows that these
programs benefit from the automatic instance restarts of AWS Batch.

\section{Discussion}\label{discussion}

The main advantage of an event-driven approach based on AWS Batch and
Lambda functions for seismic imaging in the cloud is the automated
management of computational resources by AWS. EC2 instances that are
used for carrying out heavy computations, namely for solving large-scale
wave equations, are started automatically in response to events, which
in our case are Step Functions advancing the serverless workflow to the
\texttt{ComputeGradients} state. Expensive EC2 instances are thus only
active for the duration it takes to compute one element $\mathbf{g}_i$
of the full or mini-batch gradient and they are automatically terminated
afterwards. Summing the gradients and updating the variables (i.e.~the
seismic image) is performed on cheaper Lambda functions, with billing
being again solely based on the execution time of the respective code
and the utilized memory. The cost overhead introduced by Step Functions,
SQS messages and AWS Batch is negligible compared to the cost of the EC2
instances that are required for the gradient computations, while cost
savings from spot instances and eliminating idle EC2 instances lead to
significant cost savings, as shown in our examples. With the benefits of
spot instances (factor 2-3), avoidance of idle instances and the
overhead of spinning clusters (factor 1.5-2), as well as improved
resilience, we estimate that our event-driven workflow provides cost
savings of up to an order of magnitude in comparison to using fixed
clusters of (on-demand) EC2 instances.

The second major advantage of the proposed approach is the handling of
resilience. Instead of running as a single program, our workflow is
broken up into its individual components and expressed through Step
Function states. Parallel programs based on MPI rely on not being
interrupted by hardware failures during the entire runtimetime of the
code, making this approach susceptible to resilience issues. Breaking a
seismic imaging workflow into its individual components, with each
component running as an individual (sub-) program and AWS managing their
interactions, makes the event-driven approach inherently more resilient.
On the one hand, the runtime of individual components, such as computing
a gradient or summing two arrrays, is much shorter than the runtime of
the full program, which minimizes the exposure to instance failures. On
the other hand, AWS provides built-in resilience for all services used
in the workflow. Exceptions in AWS Batch or Lambda lead to computations
being restarted, while Step Functions allow users to explicitly define
their behavior in case of exceptions, including the re-execution of
states. Similarly, messages in an SQS queue have guaranteed
at-least-once delivery, thus preventing messages from being lost.
Finally, computing an embarrassingly parallel workload with AWS Batch,
rather than as a MPI-program, provides an additional layer of
resilience, as AWS Batch processes each item from its queue separately
on an individual EC2 instance and Docker container. Instance failures
therefore only affect the respective gradient and computations are
automatically restarted by AWS Batch.

The most prominent disadvantage of the event-driven workflow is that EC2
instances have to be restarted by AWS Batch in every iteration of the
workflow. In our performance analysis, we found that the overhead of
requesting EC2 instances and starting the Docker container lies in the
range of several minutes and depends on the overall batch size and on
how many instances are requested per gradient. The more items are
submitted to a batch queue, the longer it typically takes AWS Batch to
launch the final number of instances and to process all items from the
queue in parallel. On the other hand, items that remain momentarily in
the batch queue, do not incur any cost until the respective EC2 instance
is launched. The overhead introduced by AWS Batch therefore only
increases the time-to-solution, but does not affect the cost negatively.
Due to the overhead of starting EC2 instances for individual
computations, our proposed workflow is therefore applicable if the
respective computations (e.g.~computing gradients) are both
embarrassingly parallel and take a long time to compute; ideally in the
range of hours rather than minutes. We therefore expect that the
advantages of our workflow will be even more prominent when applied to
3D seismic data sets, where computations are orders of magnitude more
expensive than in 2D. Devito provides a large amount of flexibility
regarding data and model parallelism and allows us to address the large
memory imprint of backpropagation through techniques like optimal
checkpointing or on-the-fly Fourier transforms, thereby presenting all
necessary ingredients to apply our workflow to large-scale 3D models.

Our application, as expressed through AWS Step Functions, represents the
structure of a generic gradient-based optimization algorithm and is
therefore applicable to problems other than seismic imaging and
full-waveform inversion. The design of our workflow lends itself to
problems that exhibit a certain MapReduce structure, namely they
consists of a computationally expensive, but embarrassingly parallel Map
part, as well as a computationally cheaper to compute Reduce part. On
the other hand, applications that rely on dense communications between
workers or where the quantities of interest such as gradients or
functions values are cheap to compute, are less suitable for this
specific design. For example, deep convolutional neural networks (CNNs)
exhibit mathematically a very similar problem structure to seismic
inverse problems, but forward and backward evaluations of CNNs are
typically much faster than solving forward and adjoint wave equations,
even if we consider very deep networks like ResNet \cite{He2016}.
Implementing training algorithms for CNNs as an event-driven workflow as
presented here, is therefore excessive for the problem sizes that are
currently encountered in deep learning, but might be justified in the
future if the dimensionality of neural networks continues to grow.

The event-driven workflow presented in this work was specifically
designed for AWS and takes advantage of specialized services for batch
computing or event-driven computations that are available on this
platform. However, in principle, it is possible to implement our
workflow on other cloud platforms as well, as almost all of the utilized
services have equivalent versions on Microsoft Azure or the Google Cloud
Platform (Table~\ref{t3}) \cite{MapAzure2019, MapGCP2019}. Services for
running parallel containerized workloads in the cloud, as well as
event-driven cloud functions, which are the two main components of our
workflow, are available on all platforms considered in our comparison.
Furthermore, both Microsoft Azure as well as the GCP offer similar
Python APIs as AWS for interfacing cloud services. We also speculate
that, as cloud technology matures, services between different providers
will likely grow more similar to each other. This is based on the
presumption that less advanced cloud platforms will imitate services
offered by major cloud providers in order to be competitive in the
growing cloud market.

\begin{table}[!t]
\renewcommand{\arraystretch}{1.3}
\caption{An overview how the AWS services used in our workflow map to
other cloud providers. }\label{t3}
\centering
\begin{tabular}{l l l}
\hline
Amazon Web Services & Microsoft Azure & Google Cloud \tabularnewline
\hline
Elastic Compute Cloud & Virtual Machines & Compute
Engine\tabularnewline
Simple Storage System & Blob storage & Cloud
Storage\tabularnewline
AWS Batch & Azure Batch & Pipelines\tabularnewline
Lambda Functions & Azure Functions & Cloud Functions\tabularnewline
Step Functions & Logic Apps & N/A\tabularnewline
Simple Message Queue & Queue Storage & Cloud
Pub/Sub\tabularnewline
Elastic File System & Azure Files & Cloud Filestore \tabularnewline
\hline
\end{tabular}

\end{table}

Overall, our workflow and performance evaluation demonstrate that
cost-competitive HPC in the cloud is possible, but requires a
fundamental software re-design of the corresponding application. In our
case, the implementation of an event-driven seismic imaging workflow was
possible, as we leverage Devito for expressing and solving the
underlying wave equations, which accounts for the major workload of
seismic imaging. With Devito, we are able to abstract the otherwise
complex implementation and performance optimization of wave equation
solvers and take advantage of recent advances in automatic code
generation. As Devito generates code for solving single PDEs, with the
possibility of using MPI-based domain decomposition, we are not only
able to leverage AWS Batch for the parallelization over source
experiments, but can also take advantage of AWS Batch's multi-node
functionality to shift from data to model parallelism. In contrast, many
seismic imaging codes are software monoliths, in which PDE solvers are
enmeshed with IO routines, parallelization and manual performance
optimization. Adapting codes of this form to the cloud is fundamentally
more challenging, as it is not easily possible to isolate individual
components such as a PDE solver for a single source location, while
replacing the parallelization with cloud services. This illustrates that
separation of concerns and abstract user interfaces are a prerequisite
for porting HPC codes to the cloud such that the codes are able to take
advantage of new technologies like object storage and event-driven
computations. With a domain-specific language compiler, automatic code
generation, high-throughput batch computing and serverless visual
algorithm definitions, our workflow represents a true vertical
integration of modern programming paradimgs into a framework for HPC in
the cloud.

\section{Conclusion}
Porting HPC applications to the cloud using a lift and shift approach
based on virtual clusters that emulate on-premise HPC clusters, is
problematic as the cloud cannot offer the same performance and
reliability as conventional clusters. Applications such as seismic
imaging that are computationally expensive and run for a long time, are
faced with practical challenges such as cost and resilience issues,
which prohibit the cloud from being widely adapted for HPC tasks.
However, the cloud offers a range of new technologies such as object
storage or event-driven computations, that allow to address
computational challenges in HPC in novel ways. In this work, we
demonstrate how to adapt these technologies to implement a workflow for
seismic imaging in the cloud that does not rely on a conventional
cluster, but is instead based on serverless and event-driven
computations. These tools are not only necessary to make HPC in the
cloud financially viable, but also to improve the resilience of
workflows. The code of our application is fully redesigned and uses a
variety of AWS services as building blocks for the new workflow, thus
taking advantage of AWS being responsible for resilience, job
scheduling, and resource allocations. Our performance analysis shows
that the resulting workflow exhibits competitive performance and
scalability, but most importantly minimizes idle time on EC2 instances
and cost and is inherently resilient. Our example therefore demonstrates
that successfully porting HPC applications to the cloud is possible, but
requires to carefully adapt the corresponding codes to to the new
environment. This process is heavily dependent on the specific
application and involves identifying properties of the underlying
scientific problem that can be exploited by new technologies available
in the cloud. Most importantly, it requires that codes are modular and
designed based on the principle of separation of concerns, thus making
this transition possible.

\section*{Acknowledgments}

This research was funded by the Georgia Research Alliance and the Georgia Institute of Technology.

\section*{Appendix}\label{appendix}

\subsection*{Experimental setup}

Table~\ref{t1A} lists the dimensions of the BP 2004 model and the
corresponding seismic data set that was used in our performance
analysis. Both the model and data set are publicly available from the
society of exploration geophysicists \cite{BPmodel2004}.

\begin{table}[!h]
\renewcommand{\arraystretch}{1.3}
\caption{Parameters of the BP 2004 velocity benchmark model and the
corresponding seismic data set.}\label{t1A}
\centering
\begin{tabular}{ll}
\hline
Grid dimensions & $1,911 \times 10,789$\tabularnewline
Grid spacing {[}m{]} & $6.25 \times 6.25$\tabularnewline
Domain size {[}km{]} & $11.94 \times 67.43$\tabularnewline
Number of seismic source $n_s$ & $1,348$\tabularnewline
Propagation time {[}s{]} & $12$\tabularnewline
Number of time steps & $21,889$\tabularnewline
Dimensions of each $\mathbf{d_i}$ (reshaped to 2D array) &
$2,001 \times 1,201$\tabularnewline
Dominant frequency of source {[}Hz{]} & $20$\tabularnewline
\hline
\end{tabular}

\end{table}

\subsection*{Computational resources}

Table~\ref{t2A} provides an overview of the AWS EC2 instances used in
our performance analysis, including their respective CPU architectures.
EC2 instances of a fixed instance type (such as \texttt{r5} or
\texttt{c5n}) have the same architecture for different sizes (e.g.
\texttt{2xlarge}, \texttt{4xlarge}), as those instances run on the same
hardware.

\begin{table}[!h]
\renewcommand{\arraystretch}{1.3}
\caption{Architectures of compute instances used in our performance
analysis on AWS and Optimum.}\label{t2A}
\centering
\begin{tabular}{llcc}
\hline
Instance & Intel Xeon Architecture & vCPUs & RAM
(GB)\tabularnewline
\hline
m4.4xlarge & E5-2686 v4 @ 2.30GHz & 16 &
64\tabularnewline
r5.12xlarge & Platinum 8175M @ 2.50 GHz & 48 & 
384\tabularnewline
r5.24xlarge & Platinum 8175M @ 2.50 GHz & 96 & 
768\tabularnewline
c5n.9xlarge & Platinum 8124M @ 3.00 GHz & 36 & 
384\tabularnewline
c5n.12xlarge & Platinum 8142M @ 3.00 GHz & 72 & 
768\tabularnewline
r5.metal & Platinum 8175M @ 2.50 GHz & 96 & 
768\tabularnewline
Optimum & E5-2680 v2 @ 2.80GHz & 20 & 
256\tabularnewline
\hline
\end{tabular}

\end{table}

\bibliography{manuscript_witte} 

\end{document}